\documentclass[11pt]{article}
\usepackage{graphicx}
\newlength{\bredde}
\def\slash#1{\settowidth{\bredde}{$#1$}\ifmmode\,\raisebox{.15ex}{/}
\hspace*{-\bredde} #1\else$\,\raisebox{.15ex}{/}\hspace*{-\bredde} #1$\fi}

\newcommand{\beq}{\begin{equation}}
\newcommand{\eeq}{\end{equation}}
\newcommand{\noi}{\vspace{12pt}\noindent}
\newcommand{\lG}{\raise.3ex\hbox{$\stackrel{\leftarrow}{G}$}}
\newcommand{\lU}{\raise.3ex\hbox{$\stackrel{\leftarrow}{U}$}}
\newcommand{\lP}{\raise.3ex\hbox{$\stackrel{\leftarrow}{{\cal P}}$}}
\newcommand{\leta}{\raise.3ex\hbox{$\stackrel{\leftarrow}{\eta}$}}
\newcommand{\lOmega}{\raise.3ex\hbox{$\stackrel{\leftarrow}{\Omega}$}}
\newcommand{\ldr}{\raise.3ex\hbox{$\stackrel{\leftarrow}{\delta^r}$}}

\newcommand{\beqa}{\begin{eqnarray}}
\newcommand{\eeqa}{\end{eqnarray}}
\newcommand{\n}{\newline}

\def\beqn{\begin{eqnarray}}
\def\eeqn{\end{eqnarray}}

\def\gtwid{\raise.3ex\hbox{$>$\kern-.75em\lower1ex\hbox{$\sim$}}}
\def\ltwid{\raise.3ex\hbox{$<$\kern-.75em\lower1ex\hbox{$\sim$}}}

\newcommand{\tx}[1]{\textrm{#1}}

\begin{document}

\topmargin -1.4cm
\oddsidemargin -0.8cm
\evensidemargin -0.8cm

\title{\Large{{\bf 
The Microscopic Spectral Density
of the Dirac Operator derived
from Gaussian Orthogonal and Symplectic Ensembles}}}

\vspace{1.5cm}

\author{~\\~\\
{ Christian Hilmoine } and 
{Rune Niclasen }\\~\\~\\
The Niels Bohr Institute \\ Blegdamsvej 17\\ 
DK-2100 Copenhagen, Denmark\\~\\~\\}
\maketitle
\vfill
\begin{abstract} 
The microscopic spectral correlations of the Dirac operator in
Yang-Mills theories coupled to fermions in $(2+1)$ dimensions can be
related to three universality classes of Random Matrix Theory.  In the
microscopic limit the Orthogonal Ensemble (OE) corresponds to a theory
with 2 colors and fermions in the fundamental representation and the
Symplectic Ensemble (SE) corresponds to an arbitrary number of colors
and fermions in the adjoint representation.  Using a new method of
Widom, we derive an expression for the two scalar kernels which
through quaternion determinants give all spectral correlation
functions in the Gaussian Orthogonal Ensemble (GOE) and in the the
Gaussian Symplectic Ensemble (GSE) with all fermion masses equal to
zero.  The result for the GOE is valid for an arbitrary number of
fermions while for the GSE we have results for an even number of
fermions.
\end{abstract}
\vfill
\begin{flushleft}
NBI-HE-00-19 \\
hep-th/0004081
\end{flushleft}
\thispagestyle{empty}
\newpage

\setcounter{page}{1}

\section {Introduction}
\label{intro}
Random Matrix Theory has successfully been used to extract information
about the spectral correlations of the Euclidean Dirac operator
$\slash{D}=\gamma_{\mu}(\partial_{\mu}+iA_{\mu})$ eigenvalues in the
low energy limit of Yang-Mills theories such as QCD.  In $(3+1)$
dimensions and in the low energy limit the effective Yang-Mills
partition function coincides with the partition function defined by
the chiral Random Matrix Theory ($\chi$RMT) when the so-called
microscopic limit is taken \cite{2,3,4,12}.  The low-energy
partition function describes the fermion mass dependence in the static
limit and in a finite volume of space-time $V$ \cite{3m} and is
determined alone by global symmetries.  The finite volume implies that
we restrict to the case where only the low-lying excitations (the
Goldstone modes) contribute to the field theory partition function,
while the kinetic terms of the Lagrangian are neglected.  In the
microscopic limit we look at the Dirac operator spectra on the scale
$\lambda=\mathcal{O}(V^{-1})$, which corresponds to a magnification of
the spectra in the vicinity of $\lambda=0$ on the scale $V^{-1}$.  In
case of spontaneous chiral symmetry breaking, reflected in a
condensate different from zero, $\Sigma \ne 0$, the scale $V^{-1}$
equals the average eigenvalue spacing. This follows from the
Banks-Casher relation \cite{13} $\Sigma=\pi \lim V^{-1} \rho(0)$,
where $\rho(0)$ is the spectral density of the Dirac operator evaluated
in origin and where first the thermodynamic and subsequently the chiral limit is
taken.  Thus we see that spontaneous chiral symmetry breaking is
intimately related to the spectrum of the Dirac operator in origin.

\noi 
In Random Matrix Theory no dynamical information is incorporated,
only the global symmetries of the physical system are being used and
therefore one studies the ``universal'' spectral correlations of the
eigenvalues of the considered operator.  Based on the work of
Leutwyler and Smilga \cite{3b} it has been conjectured that the
spectrum of the Dirac operator in QCD and similar theories is
universal in the microscopic limit \cite{2,6}.  This conjecture is
supported by the fact that the sum rules derived by Leutwyler and
Smilga, which involve inverse powers of the Dirac operator
eigenvalues, can be derived from RMT \cite{2,3}.  Thus, in this
limit the microscopic spectral correlation functions (of which the
microscopic spectral density is the most simple) of the Dirac operator
can be derived from a much simpler Random Matrix Theory in which only
the symmetries of the Dirac operator are the inputs.
Depending on the representation of the gauge group $SU(N_{c})$ and the
number of colors $N_{c}$, the Dirac operator belongs to one of three
universality classes, which in $\chi$RMT are represented by the
orthogonal ($\chi$OE), the unitary ($\chi$UE) and the symplectic
($\chi$SE) ensemble \cite{7}.  For each
theory the symmetries of the Dirac operator specifies one of these
universality classes.  In addition the chiral structure of the Dirac
operator in all three theories is incorporated in $\chi$RMT, requiring
a specific block structure of the matrices in the three ensembles
$\chi$UE, $\chi$OE and $\chi$SE. See \cite{3,7}  for a more detailed
discussion.  \\

\noi
Each of the three Yang-Mills theories in $(3+1)$ dimensions has an
analogue in $(2+1)$ dimensions and the effective partition function of
each theory has been showed to coincide with the partition function in
each of the three universality classes defined by  non-$\chi$
Random Matrix Theory (non-$\chi$RMT) \cite{12,3j,3j'}.
In an odd number of space-time dimensions chiral symmetry does not
exist. But in $(2+1)$ dimensions with an even number of flavors
$N_{f}$ it has been suggested  that  the spontaneous breakdown of flavor symmetry
occurs, and this is the analogue of spontaneous chiral symmetry breaking in $(3+1)$ dimensions \cite{12,piarski}.
Thus, the  argumentation  of the entire picture in $(3+1)$ dimensions has a parallel
in $(2+1)$ dimensions  \cite{12,3j,3j'}.
Here we learn again that an order parameter $\Sigma$ of the flavor symmetry breaking in $(2+1)$ dimensions
is related to the spectral density of the Dirac operator, evaluated at zero, through 
a generalization of the Banks-Casher relation. 
The lack of  chiral symmetry in $(2+1)$ dimensions
is in the three ensembles non-$\chi$OE, non-$\chi$UE and
non-$\chi$SE  reflected in the lack of the chiral block structure of
the matrices.

\noi 
Thus in  both $(3+1)$ and $(2+1)$ dimensions we have three types
of Yang-Mills theories, defined by a choice of fermion colors $N_{c}$
and a representation of the gauge group $SU(N_{c})$, and the
symmetries of the Dirac operator in each theory implies a specific
structure of the Dirac matrix.  Each theory is represented by a random
matrix ensemble: in $(3+1)$ dimensions we have the three chiral
($\chi$) ensembles and in $(2+1)$ dimensions the three non-chiral
(non-$\chi$) ensembles. 
\begin{table}[h]
\begin{center}
\begin{tabular}{||c|c|c|c|c||}\hline\hline
GAUGE GROUP & REPS. & DIRAC MATRIX  & ENSEMBLE  & $\beta$  \\ 
\hline\hline
SU($2$) &   Fundamental      & Real  &(non-)$\chi$OE & $1$ \\ \hline
SU($N_{c} \ge 3$)   & Fundamental &  Complex & (non-)$\chi$UE & $2$  \\ \hline
SU($N_{c}$)  & Adjoint & Quaternion real & (non-)$\chi$SE & $4$ \\ \hline
\end{tabular}
\end{center}
\caption{The table illustrates the classification, each
ensemble is labelled by the Dyson index $\beta=1,2,4$.
}
\label{treQCDteorier}
\end{table}

\noindent
In all three Yang-Mills theories in $(3+1)$ dimensions, as well as in
$(2+1)$ dimensions we can have  an arbitrary number
of flavors $N_{f}$ of course.
However, the coincidence of the ensembles $\beta=1$ and $\beta=4$
in $(2+1)$ dimensions with the corresponding effective field theory is only for \emph{even}
$N_{f}$. 
The relations to non-$\chi$GOE and
non-$\chi$GSE 
of the two field theories was very recently derived by  Magnea \cite{3j,3j'}.
In a theory with fermions in the fundamental representation of $SU(2)$
the following flavor symmetry breaking pattern causes the creation of the condensate $\Sigma$ \cite{3j}:
\beq
Sp(2N_{f}) \to Sp(N_{f}) \times Sp(N_{f}).
\eeq
In \cite{3j'} the flavor symmetry breaking pattern
\beq
O(N_{f}) \to O(N_{f}/2) \times O(N_{f}/2),
\eeq
is shown to be the one in a theory with $N_{c}$ arbitrary and the
adjoint representation of $SU(N_{c})$.  In this paper we want to
derive the massless microscopic spectral density of the Dirac operator
in these two field theories from the Random Matrix Theory of
non-$\chi$OE and non-$\chi$SE.

\noi
The universality 
\cite{10,3r} of the ensembles $\beta=1$, $\beta=2$
and $\beta=4$ allows the choice of a Gaussian distribution (G) in
these three ensembles, which is an advantage in view of the
calculation of the spectral correlation functions.  With the help of
orthonormal polynomials all the microscopic spectral correlation
functions have been derived in $\chi$GOE, $\chi$GUE and $\chi$GSE with
massive fermions, see review in \cite{rev1}.  
In non-$\chi$RMT, however,
only the microscopic spectral correlation functions in non-$\chi$UE
with an arbitrary number of massive fermions have been derived
\cite{12,3i,3u}.  
The two remaining universality classes are the non-$\chi$OE and
non-$\chi$SE.  
In this paper we derive the kernels
$S_{N}^{(\beta)}(x,y)$ for non-$\chi$GOE and non-$\chi$GSE with
massless fermions. They determine all massless spectral correlation
functions in these two ensembles.  Specifically we derive the massless
microscopic spectral density in the two ensembles.  A direct
verification of our results is possible through the generation of
matrices distributed according to the probability distribution in the
ensembles non-$\chi$GOE and non-$\chi$GSE.  
We also compare with the spectral sum rules recently derived in
\cite{3j,3j'}.

\noi
The traditional method to derive spectral correlation functions in
general orthogonal and symplectic ensembles with the use of
polynomials is known as Dyson's quaternion matrix method \cite{mehta}.  
In this
method the kernels $S_{N}^{(\beta)}(x,y)$, $\beta=1,4$, are
represented by special sums involving \emph{skew}-orthonormal
polynomials and the spectral correlation functions are determined by
the quaternion determinant of a quaternion matrix given by
$S_{N}^{(\beta)}(x,y)$, $\beta=1,4$.  Widom and Tracy  \cite{1d} have modified
Dysons quaternion matrix method, in the sense that the polynomials in
the relevant kernels now can be chosen arbitrary.  In \cite{1b}
the relevant kernels for the ensembles $\beta=1$ and $\beta=4$ are
given by \emph{orthonormal} polynomials.  The main content of section 3 in
this work is to provide a simple recipe for how to derive the two
kernels in an ensemble defined by a general weight function, only by
the use of orthonormal polynomials.  We avoid the actual proof \cite{1b}
and focus on the construction of a helpful machinery to derive the
needed kernels.  In section 4 we use the recipe on the
non-$\chi$Gaussian ensembles $\beta=1,4$, defined in section 2, and thus derive an
expression for the two kernels $S_{N}^{(\beta)}(x,y)$.  
Although the
equivalences between the field theory partition function and the the
partition functions in the two cases $\beta=1,4$, only are valid for
even $N_{f}$, and thus the result only has interest for even $N_{f}$ 
in Yang-Mills theory in $(2+1)$ dimensions \cite{3j,3j'}, the method gives us the
spectral correlation functions also for odd $N_{f}$ in the ensemble $\beta=1$.
In section 5 we present results  for the
microscopic spectral densities in the two ensembles and compare them with spectral sum rules and
with Monte Carlo simulations done directly on  random matrices.

\section{The non-$\chi$Gaussian ensembles}

\label{sec-chiRMT}

The non-$\chi$ random matrix model is defined by the partition
 function \cite{12,3j,3j'} 
\beq\label{Z-chi}
 \mathcal{Z}^{(\beta)}_{N_{f}}(\mathcal{M})=\int DT\
 P_{N_{f}}^{(\beta)}(T) = \int DT \ \prod_{ f=1 }^{ N_{f} }
 \det(iT+m_{f}) e^{ -\frac{N\Sigma \beta}{4} \textrm{ Tr V}
 (T^{2}) }, 
\eeq 
where $T$ is from an ensemble of hermitian $N \times N$ matrices and
the integration is taken over the Haar measure $DT$.  The Dyson index
$\beta$ has the value $\beta=1$ for the orthogonal ensemble
(non-$\chi$OE), $\beta=2$ for the unitary ensemble (non-$\chi$UE) and
$\beta=4$ for the symplectic ensemble (non-$\chi$SE), which
corresponds to real, complex and quaternion real matrix elements
respectively.  The matrix model is for a generic potential $V(T)$, but
the basic assumption of universality justify the use of a Gaussian
distribution, consistent with no additional input but the symmetries of
the system.  The condensate is $\Sigma \ne 0$, the parameter $N_{f}$
is restricted to integers and the diagonalized mass matrix
$\mathcal{M}$ is having $N_{f}$ masses in the diagonal.  The name
``non-$\chi$'' is attached due to the inclusion of the determinant in
(\ref{Z-chi}), which makes the integrand not always positive and which
makes these ensembles rather different from the usual orthogonal and
symplectic ensembles near zero.
\\
Putting all fermion masses to zero and deleting the determinant term
($N_{f}=0$) makes the ensemble equivalent to the well known
(classical) Gaussian ensemble and the replacement of the determinant
with its absolute value (or restriction to even $N_{f}$) gives the
generalized Gaussian ensemble \cite{2f}.  The matrices in the
\emph{chiral} ($\chi$) ensembles are rectangular in general and have a specific
block structure, the former a result of the incorporation of the
analogue of topological charge in the Random Matrix Theory and the
latter corresponds to the choice of a representation of the Dirac
matrix in a chiral basis.  The lack of topological charge and chiral
transformations in an odd number of space-time dimensions is in the
non-$\chi$ RMT model (\ref{Z-chi}) reflected in the quadratic matrices
with no additional constraints, but the symmetries in each of the
ensembles $\beta=1,2,4$. This is exactly what separates the non-$\chi$
ensemble from the $\chi$ ensemble.

\noi
By decomposition of the matrices $T$ one can easily transform to integration
over the $N$ eigenvalues $\{\lambda_{k}\}$ of $T$.  Choosing the Gaussian distribution
then gives  the partition function 
\beqa\label{Z-chiegen}
\mathcal{Z}^{(\beta)}_{N_{f}}(\mathcal{M}) 
= \int_{-\infty}^{\infty} \prod_{k=1}^{N} d\lambda_{k}  
\prod_{f=1}^{N_{f}/2} (\lambda_{k}^{2}+m_{f}^{2}) e^{-\frac{N\beta \Sigma}{4}\lambda_{k}^{2}}
\ \vert \Delta(\{\lambda_{k} \})  \vert^{\beta}
=
\int_{-\infty}^{\infty} \prod_{k=1}^{N} d\lambda_{k} 
\omega^{(\beta)}_{N_{f}}(\lambda_{k}) \vert
\Delta(\{\lambda_{k} \})  \vert^{\beta} ,
\eeqa                        
for even $N_{f}$, and for odd $N_{f}$  we have an extra mass $m$ which by the choice $m=0$ gives \cite{12}
\beqa\label{Z-chiegenu}
\mathcal{Z}^{(\beta)}_{N_{f}}(\mathcal{M})
= \int_{-\infty}^{\infty} \prod_{k=1}^{N} d\lambda_{k}  \lambda_{k}
\prod_{f=1}^{(N_{f}-1)/2} (\lambda_{k}^{2}+m_{f}^{2}) e^{-\frac{N\beta \Sigma}{4}\lambda_{k}^{2}}
\ \vert \Delta(\{\lambda_{k} \})  \vert^{\beta}
=
\int_{-\infty}^{\infty} \prod_{k=1}^{N} d\lambda_{k} 
\omega^{(\beta)}_{N_{f}}(\lambda_{k}) \vert
\Delta(\{\lambda_{k} \})  \vert^{\beta}, 
\eeqa   
were we have neglected all irrelevant overall factors.
Here the function
\beq\label{Van1}
\Delta(\{\lambda_{i}\})\equiv  \prod_{i<j}^{N}  (\lambda_{i}-\lambda_{j})
\eeq
is the Vandermonde determinant
 and we have defined the \emph{weight function} of the non-$\chi$ Gaussian ensemble
\beqa\label{wf}
\omega^{(\beta)}_{N_{f}}(\lambda_{k})\ \equiv \left\{
\begin{array}{ll}
\displaystyle
\prod_{f=1}^{N_{f}/2} (\lambda_{k}^{2}+m_{f}^{2}) e^{-\frac{N\beta \Sigma}{4}\lambda_{k}^{2}} &
\quad \textrm{for} \ N_{f} \ \textrm{even,} \ \nonumber\\
\displaystyle
\lambda_{k}\prod_{f=1}^{(N_{f}-1)/2} (\lambda_{k}^{2}+m_{f}^{2}) e^{-\frac{N\beta \Sigma}{4}\lambda_{k}^{2}} &
\quad \textrm{for} \  N_{f}\  \textrm{odd,} 
\end{array} \right.
\\
\eeqa
on the support $\mathcal{I}=]-\infty,\infty[$. 
In the next section we outline the general method to derive the
$m$-point spectral correlation function in a general ensemble defined by a weight function
$\omega$.  Note that, in the massless case, i.e all $m_{f}=0$, there is
no separation between the two cases of even and odd $N_{f}$.  Our goal
is to derive an important function which determines the $m$-point
spectral correlation functions in the two cases $\beta=1,4$ of the
massless non-$\chi$Gaussian ensembles.


\section{The $m$-point correlation function} 

The spectrum of the ensemble defined by the general  weight function $\omega$ on the support
$\mathcal{I}$ is described by the
\emph{m-point correlation function}
\beqa\label{RN}
R_{m}^{(\beta)}(x_{1}, \dots,x_{m})=
\frac{1}{\mathcal{Z}_{N_{f}}^{(\beta)}}\frac{N!}{(N-m)!} 
\int_{\mathcal{I}} dx_{m+1}\dots dx_{N}\ 
\prod_{j=1}^{N} \omega(x_{j}) \ \vert \Delta(\{ x_{i} \})\vert^{\beta}.
\eeqa
This function gives the probability density that $m$ of the eigenvalues,
irrespective of their ordering, are located in  infinitesimal 
neighborhoods of $x_{1},\dots,x_{m}$. 
\\
Now we will review how to derive the function $R_{m}^{(\beta)}$ with the help of 
\emph{orthonormal polynomials} in all three ensembles $\beta=1,2,4$.
Define the functions
\beq\label{varphi}
\varphi_{i}(x)\equiv p_{i}(x)\omega(x)^{1/2}, \quad i=0,1,2,\dots,
\eeq
where $\{p_{i}(x) \}$ is the sequence of  polynomials orthonormal with
respect to the weight function $\omega(x)$ on  $\mathcal{I}$. 
Thus the sequence $\{\varphi_{i}(x) \}$ 
consists of orthonormal functions on $\mathcal{I}$. 
Define the Hilbert space $\mathcal{H}$ 
 spanned by the functions 
$\varphi_{i}(x)$, $i=0,1,2,\dots, (N-1)$.
The projection operator $K$ onto the space $\mathcal{H}$
has the kernel
\beq\label{Kngen}
K_{N}(x,y)\equiv \sum_{k=0}^{N-1}\varphi_{k}(x)\varphi_{k}(y)=
\frac{a_{N}}{x-y} \Big(\varphi_{N}(x) \quad \varphi_{N-1}(x) \Big)
\left(
  \begin{array}{cc}
0 & 1 \\
1 & 0 
  \end{array} \right)
\left(
 \begin{array}{c}
\varphi_{N}(y) \\
\varphi_{N-1}(y)
  \end{array} \right),
\eeq
where the last equality follows from the important Christoffel-Darboux 
(see \cite{szego}) formula
and $a_{N}=k_{N-1}/k_{N}$,  $k_{N}$ denoting the highest coefficient in
$p_{N}(x)$. 
In the unitary ensemble ($\beta=2$) 
the use
of orthonormality and
a simple rewriting of $\vert \Delta(\{\lambda_{i}\}) \vert^{2}$ gives \cite{mehta}
\beqa\label{RN1}
R_{m}^{(2)}(x_{1}, \dots,x_{m})=\det\big[K_{N}(x_{i},x_{j}) \big]_{1 \le i,j \le m}.
\eeqa
In the orthogonal and symplectic ensembles we have an analogous result for $R_{m}^{(\beta)}$,
$\beta=1,4$. Here, however, the 
corresponding kernel $K_{N}^{(\beta)}$ is not a scalar, but a quaternion kernel 
and $R_{m}^{(\beta)}$ is represented by the quaternion determinant of quaternion matrix
$[K_{N}(x_{i},y_{j})]_{1 \le i,j \le m  }$ \cite{mehta}. 
Now, representing the
  quaternion kernels by their  $2 \times 2$ matrix representations,
whose entries all are given by one specific scalar kernel $S^{(\beta)}_{N}$,  
completes the analogous picture for the ensembles $\beta=1,4$ :
The matrix kernels
\beqa\label{KN4gen}
K_{N}^{(4)}(x,y)=\frac{1}{2}
\left( \begin{array}{cc}
S_{N }^{(4)}(x,y) & S^{(4)}D(x,y) \\
IS_{N}^{(4)}(x,y) & S_{N}^{(4)}(y,x)
\end{array} \right),
\eeqa
and
\beqa\label{KN1gen}
K_{N}^{(1)}(x,y)=
\left( \begin{array}{cc}
S_{N}^{(1)}(x,y) & S^{(1)}D(x,y) \\
IS_{N }^{(1)}(x,y)-\varepsilon(x-y) & S_{N}^{(1)}(y,x)
\end{array} \right),
\eeqa
determine the $m$-point correlation function 
\begin{equation}\label{korr.fkt.gen}
R_{m}^{(\beta)}(x_{1}, \dots,x_{m})=
\mathrm{Q}\det\big[ K_{N}^{(\beta)}(x_{i},x_{j})\big]_{1 \le i,j \le m}, \qquad \beta=1,4.
\end{equation}
Thus, the $m$-point correlation function is represented by the
quaternion determinant of a $2m \times 2m$ matrix. Here
$\varepsilon(x)=x/(2\vert x \vert)$ (the kernel of the operator
$\varepsilon$), the scalar kernel $S_{N}^{(\beta)}(x,y)$ is given by
certain sums of products involving the functions $\varphi_{n}$, and
$I$ and $D$ are integration and differentiation operators,
respectively.  The operator $S^{(\beta)}$ has kernel
$S_{N}^{(\beta)}(x,y)$, $S^{(\beta)}D(x,y)$ is the kernel of the
operator $S^{(\beta)}D$ and $IS_{N}^{(\beta)}(x,y)$ is the kernel of
$IS_{N}^{(\beta)}$. We see that the entire matrix kernel
$K_{N}^{(\beta)}(x,y)$ is given by the scalar kernel
$S_{N}^{(\beta)}(x,y)$, $\beta=1,4$.  In the classical way derived by
Dyson \cite{2a}, and later generalized by Mehta and Mahoux \cite{2b}, the
polynomials in the functions $\varphi_{n}$, and therefore in
$S^{(\beta)}_{N}(x,y)$, are chosen \emph{skew}-orthonormal with
respect to $\omega$, which through an important theorem by Dyson gives
the result (\ref{korr.fkt.gen}). In \cite{1d}, it has been shown that
any choice of a family of polynomials leads to the same matrix kernel
$K_{N}^{(\beta)}(x,y)$, $\beta=1,4$.  The choice of skew-orthonormal
polynomials leads apparently to the most simple
$K_{N}^{(\beta)}(x,y)$. But the tedious work to derive
skew-orthonormal polynomials and the lack of an relevant formula for
this analogue of the Christoffel-Darboux formula  (which is preferable
when the scaling limit $N \to \infty$ is taken) asks for a
representation of $S_{N}^{(\beta)}(x,y)$ in which the polynomials have
well known properties and the sums involving them are more easy to
deal with.  This is precisely the result of \cite{1b}.  For a weight
function $\omega$, with respect to which there exist orthonormal
polynomials, and for which the function $\omega'/\omega$ is a
\emph{rational} function, the two kernels $S_{N}^{(\beta)}(x,y)$,
$\beta=1,4$, are given by \cite{1b} :
\begin{equation}\label{3.8}
S^{(4)}_{N}(x,y)=K_{2N}(x,y)-\sum_{i>n,j=1}^{2n} [A_{0}C_{00}^{-1}C_{0}]_{ij} \psi_{i}(x)\varepsilon \psi_{j}(y),
  \end{equation}
 \begin{equation}\label{3.9}
S^{(1)}_{N}(x,y)=K_{N}(x,y)-\sum_{i \le n,j=1}^{2n}[AC(I-BAC)^{-1}]_{ji} \psi_{i}(x)\varepsilon \psi_{j}(y).
\end{equation}
Both scalar kernels $S_{N}^{(\beta)}(x,y)$, $\beta=1,4$, equal a
modified version of the scalar kernel (\ref{Kngen}) from the
corresponding unitary ensemble, plus a linear combination of $n$
functions. The number $n$ equals the sum of orders of the poles of
$\omega'/\omega$ in the extended complex plane, where every endpoint
of the support $\mathcal{I}$, where $\omega'/\omega$ is analytic, is
included as a simple pole.  Thus, for the Gaussian ensemble
\mbox{($\omega(x)=e^{-x^{2}}$, $\mathcal{I}=]-\infty,\infty[$)} we have $n=1$
because of the simple pole in $\infty$ and for the Legendre ensemble
($\omega(x)=1$, $\mathcal{I}=[-1,1]$) $n=2$ because of the two endpoints.  For the weight function $\omega$ the polynomials in (\ref{3.9})
(the ensemble $\beta=1$) are orthonormal with respect to $\omega^{2}$,
meaning we must make the shift $\omega^{1/2} \to \omega$ in
(\ref{varphi}).  In the function (\ref{3.8}), the involved 
polynomials are orthonormal
with respect to $\omega$, as in the ensemble with $\beta=2$, but here the
shift $N \to 2N$ must be taken.  The relevant Hilbert space for each
ensemble is thus
\beqa
&\beta=1 \ :\qquad \qquad & \mathcal{H}=
\textrm{span}\{\varphi_{0},\varphi_{1},\dots,\varphi_{N-1} \}, \qquad
\textrm{where} \qquad \varphi_{i}(x)=p_{i}(x)\omega(x), \nonumber\\
&\beta=4 \ :\qquad \qquad& \mathcal{H}=
\textrm{span}\{\varphi_{0},\varphi_{1},\dots,\varphi_{2N-1} \}, \qquad
\textrm{where} \qquad \varphi_{i}(x)=p_{i}(x)\omega(x)^{1/2},
\eeqa
and $p_{i}(x)$ is a polynomial of degree $i$.  The functions
$\psi_{i}(x)$, $i=1,2,\dots, 2n$, are determined from the two
functions $\varphi_{N}(x)$, $\varphi_{N-1}(x)$ ($N \to 2N $ for
$\beta=4$ and $\omega^{1/2} \to \omega$ for $\beta=1$), and the poles
of $\omega'/\omega$.  Once we have determined the $2n$ linear
independent functions, that is precisely $\psi_{i} \in \mathcal{H}$,
$i=1,2,\dots,n$, and $\psi_{i} \in \mathcal{H}^{\bot}$,
$i=(n+1),\dots,2n$, the matrices $A$, $B$ and $C$ (and $A_{0}$,
$C_{00}$ and $C_{0}$, which just are block parts of $A$ and $C$ ) can
be derived from these.  In the next section we describe the procedure
in great detail.  \\ The results (\ref{3.8}) and (\ref{3.9}) have
their origin in identities for each operator $S^{(\beta)}$,
$\beta=1,4$, which are valid for a general $\omega$, telling that
$S^{(\beta)}$ equals the projection operator $K$ onto the Hilbert
space $\mathcal{H}$, plus a correction.  This correction is in each
ensemble $\beta=1,4$ given by the operators $K$, $D$, and
$\varepsilon$, and in case the operator $[D,K]$ has finite rank, i.e
independent of $N$, the kernel of this correction constitute finitely
many terms.  When $\omega'/\omega$ is a rational function this is
precisely the case, and the choice of orthonormal polynomials with
respect to $\omega$ and $\omega^{2}$ for $\beta=4$ and $\beta=1$
respectively, leads to (\ref{3.8}) and (\ref{3.9}).
\\
In the proceeding section we construct a recipe for how to  derive  the $n$
corrections on the right hand sides of (\ref{3.8})
and (\ref{3.9}).


\subsection{Deriving the kernels $S_{N}^{(\beta)}(x_{i},x_{j})$, $\beta=1,4$, with the help of orthonormal polynomials}
For a semi-classical weight function $\omega$, i.e $\omega'/\omega$ is
a rational function, the following procedure \cite{1b} leads to the
derivation of the two associated kernels
$S_{N}^{(\beta)}(x_{i},x_{j})$, $\beta=1,4$.  We restrict
to weight functions on the form 
\beq\label{maalfkt}
\omega(x)=e^{-V(x)}, 
\eeq
fulfilling
\beq\label{restrict}
\lim_{x \to
\partial \mathcal{I} }\omega(x)=0,
\eeq
where $V(x)$ is continuously differentiable in the interior of
$\mathcal{I}$.  (For certain weight functions the condition
(\ref{restrict}) is unnecessary \cite{1c'}.)
The $n$ correction terms in each of the two ensembles are determined
by the poles of $\omega'/\omega$ and the two functions
\beqa\label{v1}
&\beta=1&:\qquad \qquad \varphi_{i}(x)= p_{i}(x) \omega(x), \quad i=
(N-1),\ N
\eeqa
\beqa\label{v2}
&\beta=4&: \qquad \qquad
\varphi_{i}(x)= p_{i}(x) \omega(x)^{\frac{1}{2}}, \quad i= (2N-1),\
2N,
\eeqa 
where the polynomials $p_{i}$ are chosen orthonormal with
respect to $\omega^{2}(x)=e^{-2V(x)}$ and $\omega(x)=e^{-V(x)}$,
respectively.  The following steps give the right hand side of
(\ref{3.8}) and (\ref{3.9}), where the specific functions given above,
belonging to each ensemble, are used.  In the ensemble $\beta=1$ we
assume that $N$ is even.  With the purpose to work with both ensembles
at the same time we make the shift $2N \to N$ for
$\beta=4$. Therefore, in the end of the procedure we must remember to
double up the matrix dimension $N \to 2 N$ for $\beta=4$.  Here is the
recipe summarizing the general method \cite{1b} for obtaining the two
kernels $S_{N}^{(\beta)}(x,y)$, $\beta=1,4$ :
\begin{itemize}
\item[(1)]
At first the main contribution to $S_{N}^{(\beta)}(x,y)$, $\beta=1,4$, is determined.
For each ensemble $\beta=1,4$, 
the kernel $K_{N}(x,y)$ is constructed from
the associated two functions, (\ref{v1}) or (\ref{v2}), through (\ref{Kngen}).
\item[(2)]
Using  the definition
\begin{eqnarray}\label{U}
U(x,y)=c^{(\beta)}\ \frac{V'(x)-V'(y)}{x-y},  \qquad c^{(\beta)}=\left\{
\begin{array}{ll}
1 \quad \textrm{for} \quad \beta=4 \nonumber\\
2 \quad \textrm{for} \quad \beta=1,
\end{array} \right.
\end{eqnarray} 
 calculate  the four functions
\begin{eqnarray}\label{Ax}
 A_{N}(x)\equiv a_{N}\int_{\mathcal{I}}dy \ \varphi_{N}(x)\varphi_{N-1}(y)U(x,y),
\end{eqnarray} 
\begin{eqnarray}\label{Bx}
B_{N}(x)\equiv a_{N}\int_{\mathcal{I}}dy \ \varphi_{N}^{2}(y)U(x,y),
\end{eqnarray} 
\begin{equation}\label{diffOrPolAC}
A(x)\equiv A_{N}(x)-\frac{c^{(\beta)}}{2}V'(x), \qquad C(x)\equiv \frac{a_{N}}{a_{N-1}}B_{N-1}(x),
\end{equation}
for each   $\beta=1,4$. 
(The $c^{(\beta=1)}=2$ comes from the factor $2$ in the exponent of the weight function $\omega^{2}$.)
The kernel of the operator $[D,K]$ is $[D,K](x,y)$ and 
it is given by 
\footnote{
This is easily derived with the help of \cite{1c'} :
\begin{equation}\label{FD1.4}
\left(
 \begin{array}{c}
\varphi'_{N}(x) \\
\varphi'_{N-1}(x)
  \end{array} \right)=
\left(
  \begin{array}{cc}
A(x) & B_{N}(x) \\
-C(x) & -A(x) 
  \end{array} \right)
\left(
 \begin{array}{c}
\varphi_{N}(x) \\
\varphi_{N-1}(x)
  \end{array} \right),
\end{equation}
and the  representation (\ref{Kngen}) of $K_{N}(x,y)$.  
} 
\begin{equation}\label{3.2}
[D,K](x,y)=
a_{N} \Big(\varphi_{N}(x) \quad \varphi_{N-1}(x) \Big)
\left(
  \begin{array}{cc}
\frac{C(x)-C(y)}{x-y} & \frac{A(x)-A(y)}{x-y} \\
\frac{A(x)-A(y)}{x-y} & \frac{B_{N}(x)-B_{N}(y)}{x-y} 
  \end{array} \right)
\left(
 \begin{array}{c}
\varphi_{N}(y) \\
\varphi_{N-1}(y)
  \end{array} \right) .
\end{equation}
Here the central matrix is vital for the derivation of the specific matrix  $A$ 
(not to be confused with the function $A(x)$) later on. 
\item[(3)]
Next, the number $n=n_{\infty}+\sum_{j=1}^{N_{p}} n_{x_{j}}$ is calculated.
The orders $\{n_{x_{j}} \}$ of the $N_{p}$ poles 
$\{x_{j} \}$ of the function $\omega'/\omega$ are added.
A pole at infinity with order $n_{\infty}$ is included as well as
the endpoints of $\mathcal{I}$  in which $\omega$ is analytic, the latter are
 counted as simple poles ($n_{x_{j}}=1$). (Notice that $(\omega^{2})'/\omega^{2}$, relevant for $\beta=1$,
has the same poles with same orders as $\omega'/\omega$.)
\item[(4)]
By writing out the power series of the rational functions $A(x)$, $B_{N}(x)$ and $C(x)$
in $(\ref{3.2})$ one can easily verify that the kernel $[D,K](x,y)$ 
can be represented by a linear combination of the $2n$ functions
\begin{equation}\label{3.3}
x^{k} \varphi_{N-1}(x), \qquad x^{k}\varphi_{N}(x), \qquad (0 \le k < n_{\infty}),
\end{equation}
\begin{equation}\label{3.4}
(x-x_{i})^{-k-1} \varphi_{N-1}(x), \qquad (x-x_{i})^{-k-1}\varphi_{N}(x), \qquad (0 \le k < n_{x_{i}}).
\end{equation}
In addition, the space, $\mathcal{G}$, spanned by these $2n$ functions has a subspace of dimension $n$ contained
in $\mathcal{H}$ and another subspace of dimension  $n$ contained in $\mathcal{H}^{\bot}$. 
Therefore, we now determine $2n$ linearly independent
 functions $\psi_{i}(x) \in \mathcal{H}$, 
$i=1,2,\dots,n$, and
$\psi_{i}(x) \in \mathcal{H}^{\bot}$, $i=(n+1),\dots,2n$.
It is easily shown that
all the functions in $\mathcal{G}$ are orthogonal to all functions
contained in
the space $\mathcal{H}/\mathcal{S}$, where 
\beqa\label{bot}
\mathcal{S}=\textrm{span}\{\varphi_{N-k},\varphi_{l} \}, \qquad \qquad 
0 < k \le n_{\infty},
\quad 0 \le k < (n-n_{\infty}),
\eeqa
is a subspace of $\mathcal{H}$ and has dimension $n$.  This implies
that the $n$ functions $\psi_{i} \in \mathcal{H}^{\bot}$ are given by
the demand $\psi_{i} \in \mathcal{S}^{\bot}$, simplifying the
derivation of the functions.  \\ We have
\begin{equation}\label{3.5}
[D,K](x,y)=\sum_{i,j=1}^{2n}A_{ij}\psi_{i}(x)  \psi_{j}(y),
\end{equation}
where the $2n \times 2n$ matrix $A$ 
now must be derived 
through (\ref{3.2})
by the specific choice of the functions $\psi_{i}(x)$, $i=1,2,\dots,2n$.
The matrix $A$ is symmetric (because $K$ is symmetric and $D$ is anti-symmetric) 
and satisfies
\begin{equation}\label{3.7}
A_{ij}=0 \quad \textrm{if}\quad i,j\le n \quad \textrm{or} \quad i,j>n,
\end{equation}
which reduces the work.
There are not other restrictions on the choice of the functions $\psi_{i}$,
but the linear independence, and the matrix $A$ is uniquely determined through
(\ref{3.2})
by the given choice.
\item[(5)]
The $2n \times 2n$ matrix $B$ defined by the inner product
\begin{equation}\label{Bij}
B_{ij}\equiv(\varepsilon \psi_{i}, \psi_{j})=(\int_{\mathcal{I}}dy\varepsilon(x-y)\psi_{i}(y)\ ,\ \psi_{j})=
\int_{\mathcal{I}}\int_{\mathcal{I}}dx \ dy \ \varepsilon(x-y)\psi_{i}(y)\psi_{j}(x),
\end{equation}
must now be  calculated.
The fact that $\varepsilon(x)=-\varepsilon(-x)$ implies 
\begin{equation}\label{-Bij}
B_{ij}=-B_{ji}, \qquad  B_{ii}=0,
\end{equation}
meaning that we only have to determine $n(2n-1)$ matrix elements.
Note that the elements are numbers depending on the parameters $N$
and $\beta$.  In general the result for $S_{N}^{(\beta)}(x,y)$ in the
scaling limit $N \to \infty$ has the highest interest.  If an
asymptotic relation for the orthonormal polynomials (and thus for
$\psi_{i}$) is well known, it is therefore relevant to examine whether
or not it is legal to put in the asymptotic relations when this limit
is taken in (\ref{Bij}).  It is certainly preferable to interchange
the integration and the limit $N \to \infty$ in (\ref{Bij}), the
justification of which is given by Lebesgue's Majorant Theorem.
The general procedure is valid for all (even) $N$, though, and in the
next steps we  work with the general matrix $B$, with no
reference to the actual calculation  of the matrix elements.

\item[(6)]
The matrices on the right hand sides of (\ref{3.8}) and
(\ref{3.9}) are given by the  $2n \times 2n$ matrices $J$ and $C$
\begin{equation}\label{J}
J_{ij}\equiv \delta_{ij}-(\delta_{ij}) {\vert}_{i,j > n} , \quad  0 \le i,j \le 2n,
\end{equation}
and
\begin{equation}\label{C}
C\equiv J+BA.
\end{equation} 
For instance for $n=1$ we have 
\beqa
J=
\left(
\begin{array}{ll}
1 & 0 \\
0 & 0
\end{array} \right).
\eeqa
The matrix $A_{0}$ is defined by the $2n \times n$ matrix produced by
deleting the last $n$ columns in $A$ and the $n \times 2n$ matrix
$C_{0}$ is defined by the matrix produced by deleting the last $n$
rows in $C$.  In addition we have the $n \times n$ matrix $C_{00}$
given by deleting the last $n$ rows and the last $n$ columns in $C$.
We need only the $n$ last rows of the matrix
\beq
 A_{0}C_{00}^{-1}C_{0},
\eeq
and the $n$ first columns of 
\beq
AC(I-BAC)^{-1}.
\eeq
\item[(7)]
The functions $\varepsilon
\psi_{i}(x)=\int_{\mathcal{I}}\varepsilon(x-y)\psi_{i}(y)$ are
ingredients in the $n$ correction terms in (\ref{3.8}) and in
(\ref{3.9}), as well as in the matrix elements $B_{ij}=(\varepsilon
\psi_{i},\psi_{j})$.  If allowed, it is highly preferable to put in
the asymptotic relation for $\psi_{i}(x)$ in the scaling limit $N \to
\infty$.
We must solve the integrals explicitly if this is not allowed.  
\item[(8)]
The results of points (1), (4), (6) and (7)
are collected to construct the right hand side of (\ref{3.8}) and (\ref{3.9}).
For $\beta=4$ we must remember the replacement $N \to 2N$ everywhere, giving  the right hand side
of (\ref{3.8}).  
\end{itemize} 


\section{The kernels $S_{N}^{(\beta)}(x_{i},x_{j})$}
In this section we use the recipe of last section to derive the two kernels
(\ref{3.8}) and (\ref{3.9})
 in the  the massless non-$\chi$ Gaussian ensemble, defined
by the choice $m_{f}=0$ in the weight function (\ref{wf}).
\\
For all $m_{f}=0$ we have the weight function 
\beqa\label{w}
\omega_{N_{f}}^{(\beta)}(x)=x^{N_{f}}e^{-cx^{2}},
\qquad \qquad \beta=1,4,
\eeqa   
on the support $\mathcal{I}=]-\infty,\infty[$, where 
$c=\frac{N \Sigma\beta}{4}$.
The \emph{spectral density} $\rho(x)$
is the $1$-point correlation function and from (\ref{RN}) we have \cite{12}
\beq\label{r-l}
\rho(-x)=(-1)^{N N_{f}}\rho(x).
\eeq
We will always take  $N$ even making the spectral density an even function, also when 
$N_{f}$ is odd. 
For \emph{even} $N_{f}$ the polynomials \cite{2f} 
\begin{equation}\label{poll}
H_{2m}^{N_{f}}(x)=(-1)^{m}( \bar{h}_{m}^{N_{f}})^{-\frac{1}{2}} c^{\frac{1+N_{f}}{4}}\mathnormal{L}_{m}^{\frac{N_{f}-1}{2}}(cx^{2}),
\qquad \bar{h}^{N_{f}}_{m}=\frac{\Gamma(\frac{N_{f}}{2}+m+\frac{1}{2})}{m!},
\end{equation}
\begin{equation}\label{polul}
\mathnormal{H}_{2n+1}^{N_{f}}(x)=(-1)^{n}(\bar{h}_{n}^{N_{f}})^{-\frac{1}{2}}c^{\frac{3+N_{f}}{4}} \  x \ \mathnormal{L}_{n}^{\frac{N_{f}+1}{2}}(cx^{2}),
\qquad \bar{h}^{N_{f}}_{n}=\frac{\Gamma(\frac{N_{f}}{2}+n+\frac{3}{2})}{n!},
\end{equation}
are orthonormal with respect to (\ref{w}) on
$\mathcal{I}=]-\infty,\infty[$.  Here the functions $L_{n}^{\alpha}(x)$
are the generalized Laguerre polynomials (which are orthonormal with
respect to $\omega(x)=x^{\alpha}e^{-x}$, $\alpha>-1$, on
$\mathcal{I}=]0,\infty[$), with the co
efficient of $x^{n}$ given by
\beq\label{kL}
k_{n}=(-1)^{n}/n!.
\eeq
We call the polynomials (\ref{poll}) and (\ref{polul})
generalized Hermite polynomials.  Notice that these polynomials
satisfy
\beq 
H^{N_{f}}_{2m}(-x)=H^{N_{f}}_{2m}(x), \quad
H^{N_{f}}_{2n+1}(-x)=-H^{N_{f}}_{2n+1}(x),
\eeq
and
\beq
H^{N_{f}}_{N}(x)=\frac{1}{x} H^{N_{f}-2}_{N+1}(x),
\eeq
the latter property is easily derived from a basic relation between
Laguerre polynomials.  Letting $\bar{k}_{i}^{N_{f}}$ be the
coefficient of the highest power in $H^{N_{f}}_{i}(x)$ (and
remembering that $N$ is even) implies
\beqa\label{aNgen}
a_{N}=\frac{ \bar{k}_{N-1} }{
\bar{k}_{N} }= \frac{ (-1)^{ \frac{N-2}{2} } \ k_{ \frac{N-2}{2} } }{
(-1)^{ \frac{N}{2} } \ k_{ \frac{N}{2} } } \frac{
(\bar{h}_{N-1}^{N_{f}})^{-\frac{1}{2}} }{ (\bar{h}_{N}^{N_{f}})^{
-\frac{1}{2} } } \frac{c^{\frac{N-2}{2}}c^{ \frac{3+N_{f}}{4} }
}{c^{\frac{N}{2}}c^{ \frac{1+N_{f}}{4} } }= \big( \frac{N}{2}
\big)^{\frac{1}{2}} c^{-\frac{1}{2}}, 
\eeqa 
where $k_{i}^{N_{f}}$ is the coefficient (\ref{kL}).  The generalized
Hermite polynomials are relevant for all three cases $\beta=1,2,4$ and
even $N_{f}$.  For $\beta=2,4$ and even $N_{f}$ we must use the
polynomials (\ref{poll}) and (\ref{polul}) which are orthonormal with
respect to (\ref{w}), but for odd $N_{f}$ we cannot use the
orthonormal polynomial method described earlier, due to the fact that
there does not exist orthonormal polynomials with respect to an odd
weight function defined on an even interval \cite{szego}.
For $\beta=1$, however, we must use the orthonormal polynomials with
respect to $\omega^{(\beta=1)}_{N_{f}}(x)^{2} =x^{2N_{f}} e^{-2c x^{2}
}$, implying that we can choose $N_{f}$ both even and odd in this case.  The
relevant polynomials are given (\ref{poll}) and (\ref{polul}) with the
replacements $N_{f} \to 2N_{f}$ and $c \to 2c$, which we will carry
out at the end of derivations.  Thus, in both cases
$\beta=1,4$ the number $N_{f}$ ($2N_{f}$) is even.  According to the
prescription of the case $\beta=4$ we will do the replacement $N \to 2N$ at
the end of the derivations.  

\noi 
Following the recipe described in
last section we now construct the kernels
$S_{N}^{(\beta)}(x_{i},x_{j})$, $\beta=1,4$, defined by the weight
function (\ref{w}).
\\
With the polynomials $H^{N_{f}}_{i}(x)$,
given by (\ref{poll}) and (\ref{polul}), and $
\omega_{N_{f}}^{(\beta)}(x)$, given by (\ref{w}), we have
\beqa\label{v22}
  \varphi_{i}(x)= H^{N_{f}}_{i}(x)\ \omega^{(\beta)}_{N_{f}}(x)^{\frac{1}{2}},
\eeqa
$i=0,1,\dots $, and $\mathcal{H}=\textrm{span}\{\varphi_{0},\varphi_{1},\dots,\varphi_{N-1} \}$.
In each ensemble $\beta=1,4$ we now 
have the kernel $K_{N}(x,y)$, through (\ref{Kngen}). 
\\
Because  $N_{f}$  is even in both ensembles we have
$\omega_{N_{f}}^{(\beta)}(x)=e^{-cx^{2}+N_{f} \ln\vert x \vert}$. 
It follows then, that
\beq\label{genGU}
U(x,y)=
\frac{V'(x)-V'(y)}{x-y}
=\frac{N_{f}}{xy}+\frac{(cx^{2})'-(cy^{2})'}{x-y}=\frac{N_{f}}{xy}+2c.
\eeq
The polynomial
$H_{N-1}^{N_{f}}(x)$ does not have a constant term, see (\ref{polul}),
and therefore the function  (\ref{Ax}) vanishes.
We have
\beqa\label{an}
\lefteqn{A_{N}(x)=
a_{N}\int_{-\infty}^{\infty}dy \ H^{N_{f}}_{N}(y) H^{N_{f}}_{N-1}(y)
\big(\frac{N_{f}}{xy}+2c \big) \ y^{N_{f}}\exp(-cy^{2})}\nonumber\\
&=&
a_{N}\frac{N_{f}}{x}\int_{-\infty}^{\infty}dy \ H^{N_{f}}_{N}(y) p_{N-2}(y)
 \ y^{N_{f}}\exp(-cy^{2})=0,
\eeqa
where $p_{N-2}(y)=H^{N_{f}}_{N-1}(y)/y$.
The function (\ref{Bx}) equals
\beq\label{eq:bn}
B_{N}(x)=
a_{N}\int_{-\infty}^{\infty}dy \ (H^{N_{f}}_{N}(y))^{2} 
\big(\frac{N_{f}}{xy}+2c \big) \ y^{N_{f}}\exp(-cy^{2}).
\eeq
It follows from the orthonormality properties that
the integral of the second term in (\ref{eq:bn}) equals $2ca_{N}$.
The integral of the first term gives
\beqa
\lefteqn{a_{N}\frac{N_{f}}{x}\int_{-\infty}^{\infty}dy \ (H^{N_{f}}_{N}(y))^{2}
 \ y^{N_{f}-1}\exp(-cy^{2})}\nonumber\\
 &=&
a_{N}\frac{N_{f}}{x} \frac{1}{N_{f}}y^{N_{f}}(H^{N_{f}}_{N}(y))^{2}\exp(-cy^{2})\vert_{-\infty}^{\infty}
-a_{N}\frac{N_{f}}{x}
\frac{1}{N_{f}}\int_{-\infty}^{\infty}dy \ y^{N_{f}}
\Big(H^{N_{f}}_{N}(y)\exp(-cy^{2}) \Big)'
 \nonumber\\
 & = &
 - a_{N}\frac{1}{x}\int_{-\infty}^{\infty}dy \ y^{N_{f}}
 \big[2H^{N_{f}}_{N}(y)H'^{N_{f}}_{N}(y)-2yc(H^{N_{f}}_{N}(y))^{2}  \big] \exp(-cy^{2})=0,
\eeqa
implying that
\beq
B_{N}(x)=2ca_{N}=\sqrt{2N}c^{\frac{1}{2}}.
\eeq
We then have the functions
\beqa\label{ABCgenG}
A(x)=-\frac{1}{2}V'(x)=-\frac{1}{2}(2cx-\frac{N_{f}}{x})=-cx+\frac{N_{f}}{2x},
\qquad \quad B_{N}(x)=2ca_{N}= \sqrt{2N}c^{\frac{1}{2}}, \nonumber\\
 C(x)=\frac{a_{N}}{a_{N-1}}B_{N-1}=2ca_{N}=B_{N}(x)=\sqrt{2N}c^{\frac{1}{2}} .
\eeqa
Now we have the  matrix in (\ref{3.2}), making it possible to determine the matrix $A$
after the choice of a number $2n$ of functions $\psi_{i}(x)$, see (\ref{3.5}). 
\\
The function
\beq
\frac{ \frac{d}{dx} \omega^{(\beta)}_{N_{f}}(x)  }{\omega^{(\beta)}_{N_{f}}(x)}=\frac{N_{f}}{x}-2cx,
\eeq
has a pole at $x=0$ and at $x=\infty$, with orders $n_{0}=1$  and $n_{\infty}=1$
respectively.
Thus we have $n=n_{\infty}+n_{0}=2$ meaning that we have to calculate two correction terms.
\\
We must find $n=2$ linearly independent functions $\psi_{1}, \psi_{2}
\in \mathcal{H}$ and $2$ linearly independent $\psi_{3}, \psi_{4} \in
\mathcal{H}^{\bot}$.  These functions must be written as linear
combinations of the functions (\ref{3.3}) and (\ref{3.4}), with only
the case $k=0$, and each $\psi_{i}$, $i=1,2,3,4$, is therefore a
linear combination of the four functions
\beqa
\varphi_{N-1}(x),  \qquad  \varphi_{N}(x), \qquad
\frac{\varphi_{N-1}(x)  }{x}, \qquad \frac{ \varphi_{N}(x)}{x}.
\eeqa
The function $\frac{\varphi_{N-1}(x)  }{x}$ is a polynomial of degree $(N-2)$ times
$\omega^{(\beta)}_{N_{f}}(x)^{1/2}$ and thus it follows immediately
that
\beqa\label{p1p2}
\psi_{1}(x)\equiv \varphi_{N-1}(x), \quad   \psi_{2}(x)\equiv \frac{\varphi_{N-1}(x)  }{x}
\in \mathcal{H}.
\eeqa
In addition, with the help of orthogonality it is easily shown that 
\beqa\label{p3p4}
\psi_{3}(x)\equiv \varphi_{N}(x), \quad   \psi_{4}(x)\equiv \frac{\varphi_{N}(x)  }{x}
\in \mathcal{H}^{\bot}.
\eeqa
The requirements are simply that $\psi_{3}\ \psi_{4} \notin \mathcal{H}$ and $\psi_{3},\psi_{4} \bot \varphi_{0}$ and
$\psi_{3},\psi_{4} \bot \varphi_{N-1}$ (see (\ref{bot})), which follows from
orthogonality.
\\
With the choices (\ref{p1p2}) and (\ref{p3p4}) we now determine the
matrix $A$ through (\ref{3.2}), (\ref{ABCgenG}) og (\ref{3.5}).
Inserting (\ref{ABCgenG}), (\ref{p1p2}) and (\ref{p3p4}), into
(\ref{3.2}) gives
\beq
[D,K](x,y)=-a_{N}
\Big[c\varphi_{N}(x)\varphi_{N-1}(y)+c\varphi_{N-1}(x)\varphi_{N}(y)+
\frac{N_{f}}{2xy}\big(\varphi_{N}(x)\varphi_{N-1}(y)+\varphi_{N-1}(x)\varphi_{N}(y) \big) \Big],
\eeq
A Comparison with  
 (\ref{3.5}) gives the  matrix elements of the
$4 \times 4$-matrix $A$.
We get
\beq\label{Agen}
A=\left(
\begin{array}{cccc}
0 & 0 & -a_{N}c& 0 \\
0 & 0 & 0 & -a_{N}\frac{N_{f}}{2}\\
-a_{N}c & 0 & 0 & 0 \\
0 & -a_{N}\frac{N_{f}}{2} & 0 & 0
\end{array}
\right)=
\left(\begin{array}{cccc}
0 & 0 & -\big( \frac{N}{2} \big)^{1/2}c^{\frac{1}{2}} & 0 \\
0 & 0 & 0 & - \big( \frac{N}{2} \big)^{1/2}c^{-\frac{1}{2}}  \frac{N_{f}}{2}\\
-\big( \frac{N}{2} \big)^{1/2}c^{\frac{1}{2}} & 0 & 0 & 0 \\
0 & - \big( \frac{N}{2} \big)^{1/2}c^{-\frac{1}{2}}  \frac{N_{f}}{2} & 0 & 0
\end{array}
\right),
\eeq
which is   symmetric and fulfills equation (\ref{3.7}). 
\\
Because of  (\ref{-Bij}) the $4 \times 4$-matrix $B$ is given by the elements
\beq
B_{12}= (\varepsilon  \psi_{1},\psi_{2}), \quad B_{13}=(\varepsilon \psi_{1},\psi_{3}),
\quad B_{14}=\quad (\varepsilon \psi_{1},\psi_{4})\quad \textrm{and}
\eeq
\beq
B_{23}= ( \varepsilon \psi_{2},\psi_{3}), \quad
B_{14}=(\varepsilon  \psi_{1},\psi_{4}),
\quad B_{34}=(\varepsilon  \psi_{3},\psi_{4}).
\eeq
The generalized Hermite polynomials $H^{N_{f}}_{i}(x)$ are either even
or odd functions and the same is true for the square root of the
weight function $\omega^{(\beta)}_{N_{f}}(x)^{1/2}$ (because $N_{f}$
is even), and thus the functions $\psi_{i}(x)$, $i=1,2,3,4$, are even
or odd.  A general calculation of the inner product gives $(\varepsilon
f,g)=0$, whenever $f$ and $g$ have same parity. From this it follows that
\beq
B_{14}= ( \varepsilon \psi_{1},\psi_{4})=B_{23}= ( \varepsilon \psi_{2},\psi_{3})=0,
\eeq
which together with (\ref{-Bij}) results in 
\beq\label{Bgen}
B=\left(
\begin{array}{cccc}
0 & ( \varepsilon \psi_{1},\psi_{2}) & ( \varepsilon \psi_{1},\psi_{3})& 0 \\
-( \varepsilon \psi_{1},\psi_{2}) & 0 & 0 & ( \varepsilon \psi_{2},\psi_{4})\\
-( \varepsilon \psi_{1},\psi_{3}) & 0 & 0 & ( \varepsilon \psi_{3},\psi_{4}) \\
0 & -( \varepsilon \psi_{2},\psi_{4}) & -( \varepsilon \psi_{3},\psi_{4}) & 0
\end{array}
\right).
\eeq
We have
\beq\label{bij}
B_{ij}=(\varepsilon \psi_{i}, \psi_{j})=
\int_{-\infty}^{\infty} dx \psi_{j}(x)\int_{-\infty}^{\infty} dy \varepsilon(x-y)\psi_{i}(y)=
\int_{-\infty}^{\infty} dx \psi_{j}(x) \frac{1}{2}\Big\{ (\int_{-\infty}^{-x}dy +\int_{-x}^{x}dy)-\int_{x}^{a}dy \Big\}\psi_{i}(y) ,
\eeq
from which it follows that
\beq\label{12b}
B_{ij}=2\int_{0}^{\infty}dx \ \psi_{j}(x)\int_{0}^{x} dy \ \psi_{i}(y),
\eeq
whenever $\psi_{i}$ is an even function.
Notice that in every element, $B_{ij}\ne 0$,  one of the functions
$\psi_{i}$ or $\psi_{j}$ is always even while the other is odd, which
together with the feature $B_{ij}=-B_{ji}$ implies that the absolute
value of all matrix elements $B_{ij}$ can be written on the form
(\ref{12b}).
Leaving these matrix elements unsolved for a while,  we now focus on the structure 
of the kernels $S_{N}^{(\beta)}(x,y)$, $\beta=1,4$.
\\
Having the matrices  $A$,  $B$ and ($n=2$) 
\beq\label{Jgen}
J=
\left(
\begin{array}{cccc}
1 & 0 & 0 & 0 \\
0 & 1 & 0 & 0 \\
0 & 0 & 0 & 0 \\
0 & 0 & 0 & 0 \\
\end{array}\right),
\qquad C=J+BA, \qquad
A_{0}=
\left(
\begin{array}{cc}
0 & 0   \\
0 & 1   \\
-(\frac{N}{2})^{1/2}c^{\frac{1}{2}} & 0   \\
0 & -(\frac{N}{2})^{1/2}c^{-\frac{1}{2}}\frac{\alpha}{2}  \\
\end{array}\right),
\eeq
and so on,
the two kernels are given by
\begin{equation}\label{hers4}
S^{(4)}_{N}(x,y)=
K_{N}(x,y)-
\psi_{3}(x)  \sum_{l=1}^{4}
[A_{0}C_{00}^{-1}C_{0}]_{3l} \ \varepsilon \psi_{l}(y)-
\psi_{4}(x)  \sum_{k=1}^{4}
[A_{0}C_{00}^{-1}C_{0}]_{4k}\  \varepsilon \psi_{k}(y),
  \end{equation}
 \begin{equation}\label{hers1}
S^{(1)}_{N}(x,y)=K_{N}(x,y)-
\psi_{1}(x)\sum_{l=1}^{4}[AC(I-BAC)^{-1}]_{l1}\ \varepsilon \psi_{l}(y)-
\psi_{2}(x)\sum_{k=1}^{4}[AC(I-BAC)^{-1}]_{k2}\ \varepsilon \psi_{k}(y),
 \end{equation}
where
\beqa\label{epsigen}
\varepsilon \psi_{i}(x)=
\frac{1}{2} \Big\{ \int_{-\infty}^{x} dy-\int_{x}^{\infty}dy \Big\}\psi_{i}(y)
=\frac{1}{2} \Big\{\int_{-\infty}^{-x}dy+
\int_{-x}^{x} dy-
\int_{x}^{\infty}dy \Big\}\psi_{i}(y),
\eeqa
 $i=1,2,3,4$.
If $\psi_{i}$ is an even function we see that
\beq\label{epsiil}
\varepsilon \psi_{i}(x)
= \int_{0}^{x}dy\psi_{i}(y),
\eeq
while if  $\psi_{i}$ is odd then
\beq\label{epsiiul}
\varepsilon \psi_{i}(x)=
-\int_{x}^{\infty}dy \psi_{i}(y).
\eeq 
\
Now we  turn to the troublesome  integrals  in the kernels. 
The explicit expressions for the four functions $\psi_{i}(x)$ are given by
$\varphi_{j}(x)=H^{(\beta)}_{j}(x) \omega^{(\beta)}_{N_{f}}(x)^{1/2}$, 
(\ref{p1p2}) and (\ref{p3p4}) and are
\beqa\label{p1p2p3p4}
\psi_{1}(x)&=&
(-1)^{\frac{N}{2}-1} \Big[ \frac{(\frac{N}{2}-1)!}{\Gamma( \frac{N}{2}+ \frac{N_{f}+1}{2} )}  \Big]^{\frac{1}{2}} \
\ c^{\frac{3+N_{f}}{4}} \
L_{\frac{N}{2}-1}^{\frac{N_{f}+1}{2}}(cx^{2}) x^{\frac{N_{f}}{2}+1}
e^{-c\frac{x^{2}}{2}},\nonumber\\
\psi_{2}(x)&=&\frac{\psi_{1}(x)}{x},
\nonumber\\
\psi_{3}(x)&=&
(-1)^{\frac{N}{2}} \Big[ \frac{(\frac{N}{2})!}{\Gamma( \frac{N}{2}+ \frac{N_{f}+1}{2} )}  \Big]^{\frac{1}{2}} \
 c^{\frac{1+N_{f}}{4}} \ L_{\frac{N}{2}}^{\frac{N_{f}-1}{2}}(cx^{2}) x^{\frac{N_{f}}{2}}
e^{-c\frac{x^{2}}{2}}, \nonumber\\
\psi_{4}(x)&=&\frac{\psi_{3}(x)}{x}.
\eeqa
For relatively large $N$ it is clear 
that an actual calculation of the two
kernels $S_{N}^{(\beta)}(x,y)$ becomes quite involved due to the presence of
the integrals (\ref{12b}), (\ref{epsiil}) and (\ref{epsiiul}).
Actually, it is the microscopic limit of the kernel
\beq\label{lim}
S_{N}^{(\beta)}(x,y)_{s}\equiv 
\lim_{N \to \infty}\frac{1}{N \Sigma}S_{N}^{(\beta)}(\frac{x}{N \Sigma},\frac{y}{N \Sigma}), \qquad \beta=1,4, 
\eeq
giving all the microscopic correlations, which has physical interest.
Therefore it is necessary to determine whether it is allowed to substitute
an asymptotic relation for the polynomials in the integrals when the
microscopic limit is taken. If the answer to this is positive, we
can get rid of the $N$ dependence and derive a closed analytical
expression for the microscopic kernel (\ref{lim}).  With the help of
Lebesgue's Majorant Theorem this question is pursued and answered in
Appendix A.  For (\ref{epsiil}), of course, the use of an asymptotic
relation for $\psi_{i}$ is allowed when the microscopic limit is
taken, because the integral is finite for all $N$.  We must check that
the second equality in the following equation
\beq\label{h20}
\lim_{N \to \infty}N^{\lambda}B_{ij}=
\lim_{N \to \infty}N^{\lambda}  2\int_{0}^{\infty}dx \ \psi_{j}(x)\int_{0}^{x} dy \ \psi_{i}(y)
=
 2\int_{0}^{\infty}dx \lim_{N \to \infty} N^{\lambda}\ \psi_{j}(x)\int_{0}^{x} dy \ \psi_{i}(y),
\eeq
as well as the last equality in ( see (\ref{epsiiul}))
\beqa\label{epsiiullim}
\lim_{N \to \infty}N^{\delta} \varepsilon \psi_{i}(\frac{x}{N})=
- \lim_{N \to \infty}N^{\delta}
\int_{\frac{x}{N}  }^{\infty}dy \psi_{i}(y)=
- \lim_{N \to \infty}N^{\delta}\frac{1}{N}
\int_{x}^{\infty}du \psi_{i}(\frac{u}{N})=
-\int_{x}^{\infty}du   \lim_{N \to \infty}N^{\delta-1}   \psi_{i}(\frac{u}{N}),\nonumber\\
\eeqa
are valid. Here we assume that the power of $N^{\lambda}$ and
$N^{\delta}$ have the exact values needed to make the number
$N^{\lambda}B_{ij}$ and the function $N^{\delta-1} \psi_{i}(x/N)$
convergent for $N \to \infty$. This is a consequence of the assumption
that the microscopic kernel (\ref{lim}) is finite.
\\
Now let us examine (\ref{h20}) for $i=1$ and $j=2$. After
two substitutions we have
\beqa\label{der1}
B_{12} =
C_{N}
\int_{0}^{\infty}dz \ \ L_{\frac{N}{2}-1}^{\frac{N_{f}+1}{2}}(2z)
z^{\frac{1}{2}(\frac{N_{f}}{2}-1)    }
e^{-z} \int_{0}^{z}du\ 
 \ L_{\frac{N}{2}-1}^{\frac{N_{f}+1}{2}}(2u)
 u^{      \frac{N_{f}}{4}       }
e^{-u},
\eeqa
where all irrelevant factors have been collected in $C_{N}$.  (We
remember that (\ref{12b}) only holds when $\psi_{i}$ is an even
function, meaning that  $N_{f}/2$ is odd in case of $i=1$  (see
(\ref{p1p2p3p4}) ).  But the fact $B_{ij}=-B_{ji}$ implies that all
$B_{ij}$ can be represented on the form (\ref{12b}), when we
keep track of the sign.) Using the result of Appendix A.2 we
find that (\ref{h20}) does not hold for $B_{12}$.
From (\ref{2appen1}) and point (2) at the end of A.2, this is a consequence
of
\beq
-\Big(\frac{N_{f}+1}{4}+\frac{1}{4}-\frac{N_{f}}{4}\Big)=-\frac{1}{2}\ge -1,
\eeq
and
\beq
-\Big(\frac{N_{f}+1}{4}+\frac{1}{4}-(\frac{N_{f}}{4}-\frac{1}{2}) \Big)=-1 \ge -1.
\eeq
A similar treatment of the explicit expressions for the three other
matrix elements $B_{13}$, $B_{24}$ and $B_{34}$ gives that the
corresponding equation (\ref{h20}) does \emph{not} hold for any of
them. An undesirable consequence is that we have to solve all four
integrals $B_{12}$, $B_{13}$, $B_{24}$ and $B_{34}$ before enlarging
$N$ in the kernels (\ref{hers4}) and (\ref{hers1}).
\\
Turning to the question of the validity of the last equality in
(\ref{epsiiullim}) we must use the result of Appendix A.1.  Consider for
instance $\psi_{1}$ and assume that this function is odd, that is
$\big((N_{f}/2)+1\big)$ is odd. Then from (\ref{p1p2p3p4}) and
(\ref{epsiiul}) we have
\beqa\label{h9}
\varepsilon \psi_{1}(\frac{x}{N})&=&
-\frac{1}{N}
\int_{x}^{\infty}du \ \psi_{1}(\frac{u}{N})
\propto
-\frac{1}{N}  \int_{x}^{\infty}du \
L_{\frac{N}{2}-1}^{\frac{N_{f}+1}{2}}(c\frac{u^{2}}{N^{2}})
\Big(\frac{u}{N}   \Big)^{\frac{N_{f}}{2}+1}
e^{-c\frac{u^{2}}{2N^{2}}}\nonumber\\
&\propto&
-\int_{x}^{\infty}dt \
L_{\frac{N}{2}-1}^{\frac{N_{f}+1}{2}}(2t) t^{\frac{N_{f}}{4}}
e^{-t},
\eeqa
where all irrelevant factors have been skipped.  From (\ref{1appen2})
and (\ref{betb}) it now follows that (\ref{epsiiullim}) is \emph{not}
valid for $i=1$.  This is caused by the fact
\beq
\frac{N_{f}+1}{4}-\frac{N_{f}}{4}=\frac{1}{4} \le \frac{3}{4}.
\eeq
Considering the functions $\varepsilon \psi_{i}(x)$, $i=2,3,4$, with
$\psi_{i}(x)$ odd (which of course is not possible for all functions
at the same time), an analogue argumentation gives the same
conclusion.  When the function $\psi_{i}(x)$, $i=1,2,3,4$, is odd we
therefore have to solve the integral $\varepsilon \psi_{i}(x)$ to make
an exact calculation of the kernels (\ref{hers1}) and (\ref{hers4})
possible (within reasonable time) for large $N$.

\noi 
We have learned that for finite $N$ a calculation of
$B_{12}$, $B_{13}$, $B_{24}$ and $B_{34}$ and an analytical
expression for the four functions $\varepsilon \psi_{i}(x)$,
$i=1,2,3,4$, are needed to make the results (\ref{hers1}) and
(\ref{hers4}) usable. This requirement has been met in Appendix B.


\noi
In Appendix B.1 we derive an expression for the function
\beq
\mathcal{E}_{[\bar{\alpha},\bar{\beta},n]}(x) \equiv \varepsilon \
 L_{n}^{\bar{\alpha}}(x^{2})
x^{\bar{\beta}} e^{-\frac{x^{2}}{2}},
\eeq
for integers $n, \bar{\beta}$ and $\alpha>-1$.  The functions
$\varepsilon \psi_{i}(x)$, $i=1,2,3,4$, belong to this class of
functions and it is easily derived that (see (\ref{p1p2p3p4}))
\beq\label{epsiap1}
\varepsilon \psi_{1}(x)=
k_{1}c^{-\frac{1}{4}}
\mathcal{E}_{\big[\frac{N_{f}+1}{2},\ \frac{N_{f}}{2}+1,\ \frac{N}{2}-1 \big]}(c^{\frac{1}{2}}x),
\eeq
\beq\label{epsiap2}
\varepsilon \psi_{2}(x)=k_{2} c^{\frac{1}{4}}
\mathcal{E}_{\big[\frac{N_{f}+1}{2},\ \frac{N_{f}}{2},\ \frac{N}{2}-1 \big]}(c^{\frac{1}{2}}x),
\eeq
\beq\label{epsiap3}
\varepsilon \psi_{3}(x)=k_{3}  c^{-\frac{1}{4}}
\mathcal{E}_{\big[\frac{N_{f}-1}{2},\ \frac{N_{f}}{2},\ \frac{N}{2}  \big]}(c^{\frac{1}{2}}x),
\eeq
\beq\label{epsiap4}
\varepsilon \psi_{4}(x)=k_{4} c^{\frac{1}{4}}
\mathcal{E}_{\big[\frac{N_{f}-1}{2},\ \frac{N_{f}}{2}-1, \ \frac{N}{2} \big]}(c^{\frac{1}{2}}x),
\eeq
with the coefficients
\beqa\label{k}
k_{1}&=&
(-1)^{\frac{N}{2}-1} \Big[ \frac{(\frac{N}{2}-1)!}{\Gamma( \frac{N}{2}+ \frac{N_{f}+1}{2} )}  \Big]^{\frac{1}{2}},\nonumber\\
k_{2}&=&k_{1},\nonumber\\
k_{3}&=&
(-1)^{\frac{N}{2}} \Big[ \frac{(\frac{N}{2})!}{\Gamma( \frac{N}{2}+ \frac{N_{f}+1}{2} )}  \Big]^{\frac{1}{2}},\nonumber\\
k_{4}&=&k_{3}.
\eeqa
For each $\psi_{i}$ the
associated function $\mathcal{E}$ above is given by (\ref{eLulige}).
  From (\ref{p1p2p3p4}) it is clear that if $N_{f}/2$
is even, then $\psi_{1}$ and $\psi_{4}$ are odd functions and
$\psi_{2}$ and $\psi_{3}$ are even functions.  For odd $N_{f}/2$, of
course, the situation is the other way around.  Thus for each kernel
(\ref{hers1}) and (\ref{hers4}) our derived expression will split into
two parts, one for even $N_{f}/2$ and one for odd $N_{f}/2$.
\\
In Appendix B.2 
the result for the function  $\mathcal{E}$ is used to calculate
\beq\label{der2}
\mathcal{B}_{ij}\equiv(\varepsilon \phi_{i},\phi_{j})=
\int_{-\infty}^{\infty}dx \ \phi_{j}(x) \ \int_{-\infty}^{\infty}dy \
\varepsilon(x-y) \phi_{i}(y)
\eeq
where
\beq\label{der3'}
\phi_{j}(x)=L_{m}^{\alpha}(x^{2}) x^{\beta} e^{-\frac{x^{2}}{2}}\quad \textrm{ og} \quad
\phi_{i}(x)= L_{n}^{\bar{\alpha}}(x^{2})
x^{\bar{\beta}} e^{-\frac{x^{2}}{2}} ,
\eeq
and $n,m,\beta,\bar{\beta}$ are integers, fulfilling certain
conditions.  We have always an odd $\bar{\beta}$, meaning that
$\phi_{i}$ is odd.  By keeping track of a sign we can always, with the
help of $B_{ij}=-B_{ji}$ as described earlier, secure that the
function in the inner integral of the double integral
$B_{ij}=(\varepsilon \psi_{i},\psi_{j})$ is odd (or even).  Thus the
four relevant numbers $B_{ij}$ are contained in the derived
expressions for (\ref{der2}) in Appendix B.2.
\\
Putting the functions (\ref{p1p2p3p4}) into (\ref{der2}), and
remembering the mentioned possible change of sign when we secure that
the odd function comes in the inner integral of $B_{ij}$, gives the
following result for the four matrix elements of $B$ :
\beq\label{b12gen}
B_{12}=(-1)^{\frac{N_{f}}{2}}k_{1}^{2} \mathcal{B}_{12},
\eeq
\beq\label{b13gen}
B_{13}=(-1)^{\frac{N_{f}}{2}}c^{-\frac{1}{2}}k_{1}k_{3} \mathcal{B}_{13},
\eeq
\beq\label{b24gen}
B_{24}=(-1)^{\frac{N_{f}}{2}+1}c^{\frac{1}{2}}  k_{2}k_{4} \mathcal{B}_{24},
\eeq
\beq\label{b34gen}
B_{34}=(-1)^{\frac{N_{f}}{2}+1} (k_{3})^{2} \mathcal{B}_{34},
\eeq
where the numbers $\mathcal{B}_{ij}$  are found with the help of  table
\ref{tab1} and \ref{tab2}.

\begin{table}[h]
\begin{center}
\begin{tabular}{||c|c|c||c||}\hline\hline
even $\frac{N_{f}}{2}$ &  \textrm{given by eq.}  &  \textrm{with parameters}
   \\ \hline \hline
$\mathcal{B}_{12}$ &
\textrm{ (\ref{bereg1B}), \emph{without} the $(+)$-term } & 
\small{$\alpha,\bar{\alpha} = \frac{N_{f}+1}{2},
\ \beta=\frac{N_{f}}{2},\ \bar{\beta}=\frac{N_{f}}{2}+1, 
\ n=m=\frac{N}{2}-1$} \\ \hline
$\mathcal{B}_{13}$ &
\textrm{ (\ref{bereg1B}), \emph{without} the $(+)$-term }&
\small{$\alpha = \frac{N_{f}-1}{2}, \ \bar{\alpha} = \frac{N_{f}+1}{2},
\ \beta=\frac{N_{f}}{2},\ \bar{\beta}=\frac{N_{f}}{2}+1, 
\ m=\frac{N}{2}, \  n=\frac{N}{2}-1$} \\ \hline
$\mathcal{B}_{24}$ &
\textrm{ (\ref{bereg2}) }&
\small{$\alpha = \frac{N_{f}+1}{2}, \ \bar{\alpha} = \frac{N_{f}-1}{2},
\ \beta=\frac{N_{f}}{2},\ \bar{\beta}=\frac{N_{f}}{2}-1, 
\ m=\frac{N}{2}-1, \  n=\frac{N}{2}$} \\ \hline
$\mathcal{B}_{34}$ &
\textrm{ (\ref{bereg1B}), \emph{without} the $(+)$-term}&
\small{$\alpha = \bar{\alpha} = \frac{N_{f}-1}{2},
\ \beta=\frac{N_{f}}{2}-1,\ \bar{\beta}=\frac{N_{f}}{2}, 
\ m= n=\frac{N}{2}$} \\ \hline
\end{tabular}
\caption{A recipe giving $\mathcal{B}_{ij}$ for \emph{even} $N_{f}/2$
.}
\label{tab1}
\end{center}
\end{table}

\begin{table}[h]
\begin{center}
\begin{tabular}{||c|c|c||c||}\hline\hline
odd $\frac{N_{f}}{2}$ &  \textrm{given by eq.}  &  \textrm{with parameters}
   \\ \hline \hline
$\mathcal{B}_{12}$ &
\textrm{ (\ref{bereg1B}), \emph{without} the $(+)$-term } & 
\small{$\alpha,\bar{\alpha} = \frac{N_{f}+1}{2},
\ \beta=\frac{N_{f}}{2}+1,\ \bar{\beta}=\frac{N_{f}}{2}, 
\ n=m=\frac{N}{2}-1$} \\ \hline
$\mathcal{B}_{13}$ &
\textrm{ (\ref{bereg1B}), \emph{including} the $(+)$-term }&
\small{$\alpha = \frac{N_{f}+1}{2}, \ \bar{\alpha} = \frac{N_{f}-1}{2},
\ \beta=\frac{N_{f}}{2}+1,\ \bar{\beta}=\frac{N_{f}}{2}, 
\ m=\frac{N}{2}-1, \  n=\frac{N}{2}$} \\ \hline
$\mathcal{B}_{24}$ &
\textrm{ (\ref{bereg1B}), \emph{without} the $(+)$-term }&
\small{$\alpha = \frac{N_{f}-1}{2}, \ \bar{\alpha} = \frac{N_{f}+1}{2},
\ \beta=\frac{N_{f}}{2}-1,\ \bar{\beta}=\frac{N_{f}}{2}, 
\ m=\frac{N}{2}, \  n=\frac{N}{2}-1$} \\ \hline
$\mathcal{B}_{34}$ &
\textrm{ (\ref{bereg1B}), \emph{without} the $(+)$-term}&
\small{$\alpha = \bar{\alpha} = \frac{N_{f}-1}{2},
\ \beta=\frac{N_{f}}{2},\ \bar{\beta}=\frac{N_{f}}{2}-1, 
\ m= n=\frac{N}{2}$} \\ \hline
\end{tabular}
\caption{A recipe giving $\mathcal{B}_{ij}$ for \emph{odd} $N_{f}/2$.}
\label{tab2}
\end{center}
\end{table}
\pagebreak
\noindent
Having calculated the four matrix elements of $B$ and the four functions $\varepsilon \psi_{i}$
we can now construct the two kernels (\ref{hers1}) and (\ref{hers4}) of the non-$\chi$Gaussian
ensemble.
The proceeding points summarize the construction of the kernels :
\begin{itemize}
\item[(1)]
Through equation (\ref{Kngen}),
the kernel $K_{N}(x,y)$ is given by (\ref{v22}) with the associated comments.
\item[(2)]
Assume that $(\frac{N_{f}}{2})$ is \emph{even} (odd).  Then it follows
from (\ref{p1p2p3p4}), that $\psi_{1}$ and $\psi_{4}$ are odd (even),
and $\psi_{2}$ and $\psi_{3}$ are even (odd).  
The functions
$\varepsilon \psi_{1}$, $\varepsilon \psi_{2}$,
$\varepsilon \psi_{3}$ and $\varepsilon \psi_{4}$
 are given by
(\ref{epsiap1}), (\ref{epsiap2}),
(\ref{epsiap3}) and (\ref{epsiap4}),
 respectively, with $\mathcal{E}$
given by (\ref{eLulige}) both for even and  odd  $\psi_{i}$.
\item[(3)]
The matrix elements $B_{12}$, $B_{13}$, $B_{24}$ and $B_{34}$ depend
on whether $N_{f}/2$ is even or odd.  These are given by
(\ref{b12gen}), (\ref{b13gen}), (\ref{b24gen}) and (\ref{b34gen}),
with $\mathcal{B}_{ij}$ given by tables \ref{tab1} and \ref{tab2}.
Matrix $B$ is given by (\ref{Bgen}).
\item[(4)]
The matrix
\beq
C=J+BA
\eeq
 is given by (\ref{Agen}), (\ref{Bgen}) and (\ref{Jgen}).
We need the matrices
\beqa
A_{0}C_{00}^{-1}C_{0} &\textrm{for}& \quad \beta=4,
\quad \textrm{and} \nonumber\\
\quad AC(I-BAC)^{-1} \quad &\textrm{for}& \quad \beta=1.
\eeqa 
\item[(5)]

Before putting all pieces together in the construction of (\ref{hers1})
and (\ref{hers4}) we remember  the replacements
\beqa\label{sub1}
&\beta=1&  : \qquad c \to 2c \qquad \textrm{and}  \qquad N_{f} \to 2 N_{f}, \nonumber\\
&\beta=4& : \qquad N \to 2N,
\eeqa
in all parts above and in the functions (\ref{p1p2p3p4}). 
In  the two ensembles $\beta=1,4$ we have $c=\frac{N \Sigma^{2}\beta}{4}$.
\end{itemize}
The five points above leads to two formulas for each $S_{N}^{(\beta)}(x,y)$, $\beta=1,4$ :
one for even  $N_{f}$ and one for odd $N_{f}$ for $\beta=1$, and
one for even  $N_{f}/2$ and one for odd $N_{f}/2$ for $\beta=4$.

\section{The microscopic spectral density}

In the microscopic limit the spectral correlations between the eigenvalues of
the Dirac operator in the two Yang-Mills theories mentioned in the
introduction can be derived from the corresponding symplectic or
orthogonal non-$\chi$Gaussian ensembles. In this section we present
our results for our numerical expressions for the microscopic spectral density.

\noi
From (\ref{korr.fkt.gen}) it follows that the $1$-point correlation function, that is the spectral density, is given by
\beqa
\rho^{(\beta)}(x)\equiv R^{(\beta)}_{N}(x)= S_{N}^{(\beta)}(x,x), \qquad \beta=1,4.
\eeqa
The machinery from the last section gives both $S_{N}^{(\beta)}(x,y)$, $\beta=1,4$,
and we therefore have in principle   all  correlation functions and 
especially the microscopic  spectral density in the cases $\beta=1$ and $\beta=4$
of the non-$\chi$Gaussian ensemble.
The \emph{microscopic} spectral density is defined by
\beq\label{msd}
\rho_{s}^{(\beta)}(x)\equiv \lim_{N \to \infty} \frac{1}{N}\rho^{(\beta)}(\frac{x}{N}),\qquad \beta=1,4,
\eeq
and this is of physical interest. 
We have not  yet extracted the $N$ dependence of our
results for the  $B_{ij}$ (derived in
appendix B.2) and for the functions $\varepsilon \psi_{i}$ (derived in
appendix B.1) in the microscopic limit   
and we  therefore  have not  an exact analytical result for
$\rho_{s}^{(\beta)}(x)$, $\beta=1,4$.  However, we can easily present
plots of the scaled spectral density
\beq\label{ssd}
\rho_{N}^{(\beta)}(x)\equiv  \frac{1}{N}\rho^{(\beta)}(\frac{x}{N})=
\frac{1}{N}S_{N}^{(\beta)}(\frac{x}{N},\frac{x}{N}),\qquad \beta=1,4,
\eeq
which, of course, converges towards  $\rho_{s}^{(\beta)}(x)$ for large $N$. 

\begin{figure}[!htbp]
\begin{tabular}{c c}
\includegraphics[scale=0.9]{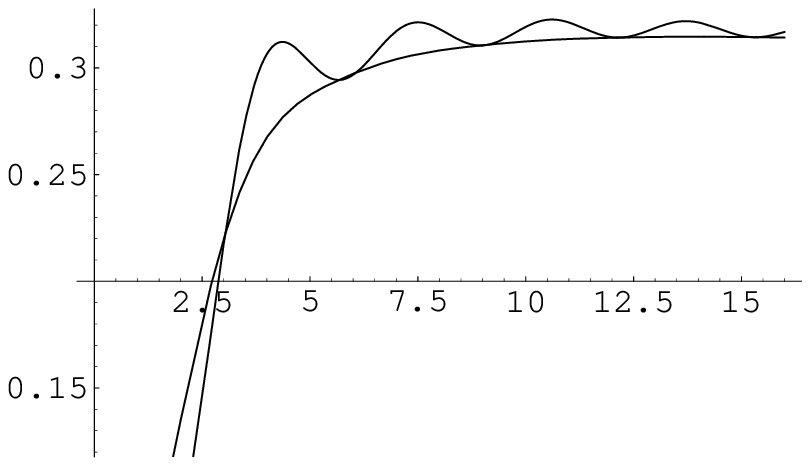} &  \includegraphics[scale=0.9]{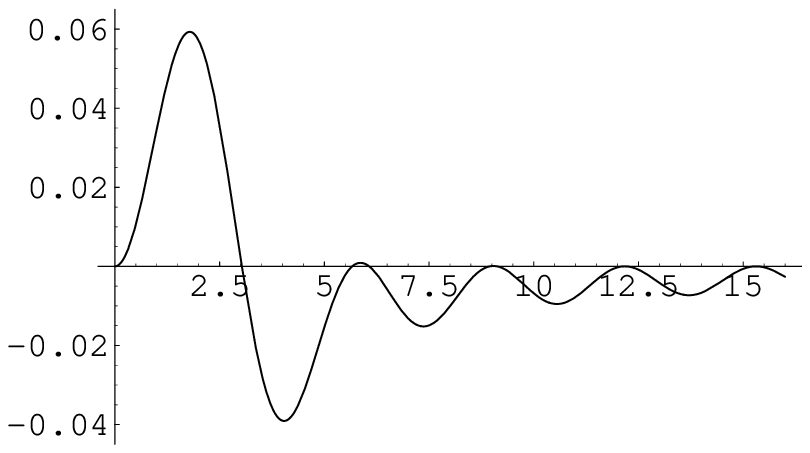} \\
\includegraphics[scale=0.9]{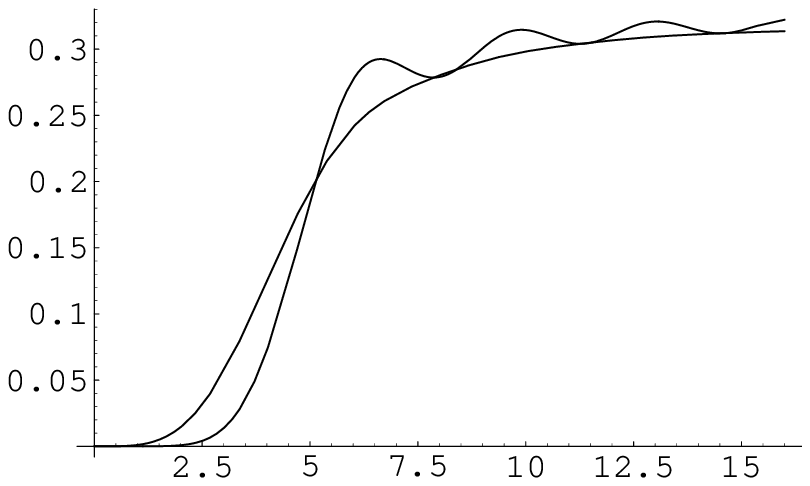} &  \includegraphics[scale=0.9]{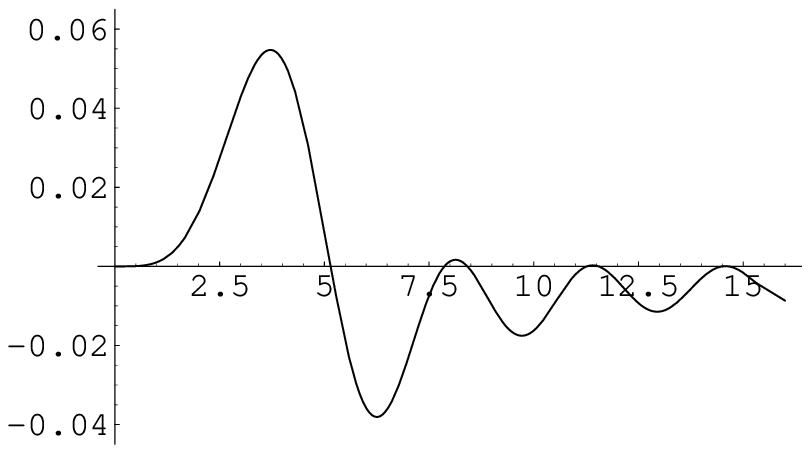} 
\end{tabular}
\caption{In non-$\chi$GOE we have  $\rho_{N}^{(1)}(x)=\rho^{(2)}_{N}(x)+\rho_{N \ \tx{corr.}}^{(1)}(x)$, where
$\rho^{(2)}_{N}(x)$ is from the associated unitary ensemble  (given by orthonormal polynomials 
with respect to $\omega_{N_{f}}^{(1)}(x)^{2}$) and  $\rho_{N \ \tx{corr.}}^{(1)}(x)$ is the scaled correction
derived in last section. 
In left column the oscillating  term $\rho^{(2)}_{40}(x)$ and the entire $\rho_{40}^{(1)}(x)$ are plotted together 
for $N=40$ and the two values $N_{f}=2$ (above) and $N_{f}=4$ (under). 
The corresponding correction terms $\rho_{40 \ \tx{corr.}}^{(1)}(x)$ are plotted in right column.
We observe that
the flat spectrum of non-$\chi$GOE is exactly a result of  $\rho_{40 \ \tx{corr.}}^{(1)}(x)$, cutting of
the peaks of $\rho_{40}^{(2)}(x)$.
In addition the curve of $\rho_{40}^{(1)}(x)$ moves away from zero
when $N_{f}$ is enlarged, due to the presence of the determinant function
in the probability distribution (see (\ref{Z-chi})).} 
\label{fig:6}
\end{figure}

\begin{figure}[!htbp]
\begin{center}
\begin{tabular}{c c}
\includegraphics[scale=0.9]{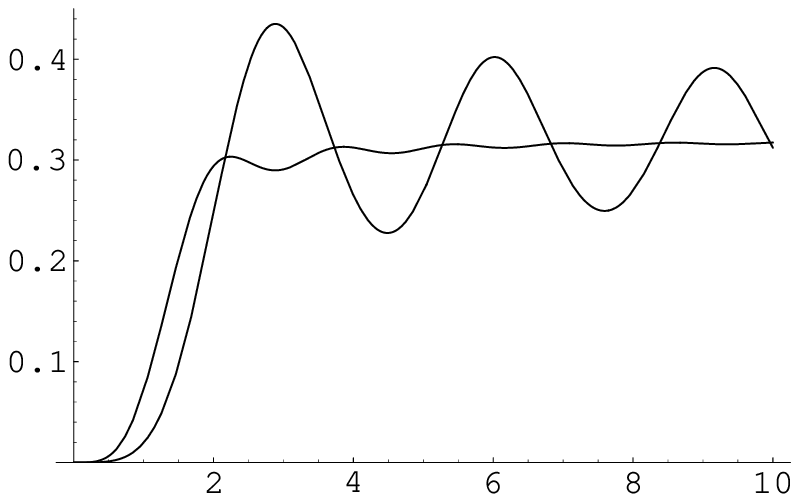} & 
\includegraphics[scale=0.9]{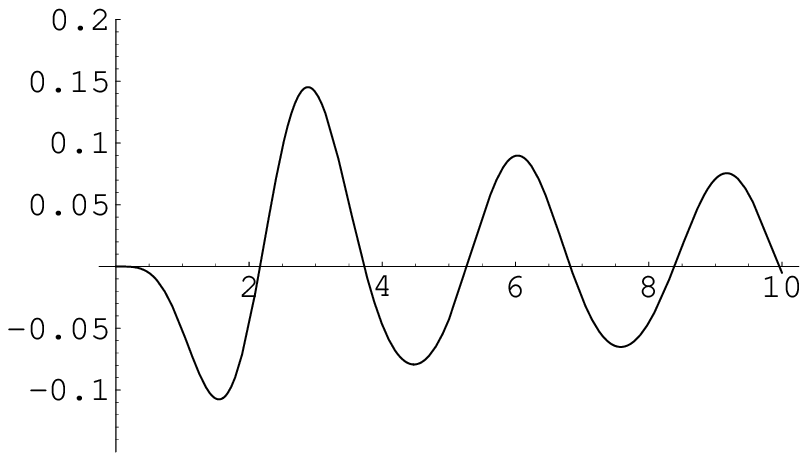} \\
\end{tabular}
\end{center}
\caption{In non-$\chi$GSE we have  $\rho_{N}^{(4)}(x)=\rho^{(2)}_{N}(x)+\rho_{N \ \tx{corr.}}^{(4)}(x)$, where
$\rho^{(2)}_{N}(x)$ is from the associated unitary ensemble  (given by orthonormal polynomials 
with respect to $\omega_{N_{f}}^{(4)}(x)$ and the replacement $N \to 2N$) 
and  $\rho_{N \ \tx{corr.}}^{(4)}(x)$ is the scaled correction
derived in last section. 
On the left the term $\rho^{(2)}_{40}(x)$ and the entire $\rho_{40}^{(1)}(x)$ are plotted together 
for $N=40$ and the value  $N_{f}=4$. 
The corresponding correction term $\rho_{40 \ \tx{corr.}}^{(4)}(x)$ is
plotted on the right.
We observe that  $\rho_{40 \ \tx{corr.}}^{(4)}(x)$ gives a highly oscillating $\rho_{40}^{(4)}(x)$
(a rigid spectrum). 
 }
\label{fig:7}
\end{figure}

\noi
Due to the presence of the Vandermonde determinant in the probability 
distribution one expects that the spectrum
becomes more rigid when the parameter $\beta$ is increased (see (\ref{RN})).  
Thus we
naturally expect  the scaled spectral density $\rho_{N}^{(1)}(x)$ in
the ensemble non-$\chi$GOE  to be flat compared to
$\rho_{N}^{(2)}$ in non-$\chi$GUE, and that
$\rho_{N}^{(4)}$ in non-$\chi$GSE  is the most
oscillating function of the three of them.  These features are of
course valid for all values of $N$. \\ 
In both ensembles $\beta=1,4$ the kernel $S_{N}^{(\beta)}(x,y)$
is equal to a kernel $K_{(2)N}(x,y)$ from the unitary ensemble $\beta=2$
(remembering the modifications associated to each value  of $\beta=1,4$),
plus $n$ corrections,  in general  (see (\ref{3.8}) and (\ref{3.9})).
Thus in the two non-$\chi$ ensembles 
we have $\rho_{N}^{(\beta)}(x)=\rho^{(2)}_{N}(x)+\rho_{N \ \textrm{corr.}}^{(\beta)}(x)$, $\beta=1,4$,
where
$\rho^{(2)}_{N}(x)=(N^{-1})K_{(2)N}(N^{-1}x,N^{-1}x)$ and $\rho_{N\ \textrm{corr}}^{(\beta)}(x)$ constitute the $2$ correction terms.
When having calculated the matrix $B$ and the four functions $\varepsilon \psi_{i}(x)$ both scaled corrections $\rho_{N \ \textrm{corr.}}^{(\beta)}(x)$, $\beta=1,4$, 
are given by the two last terms on the right hand side of 
(\ref{hers4}) and (\ref{hers1}). 
The contribution from the 
scaled spectral density  
$\rho^{(2)}_{N}(x)$ 
is an oscillating function in both ensembles $\beta=1,4$,
 and the  flat spectrum of $\rho^{(1)}_{N}(x)$ in 
the ensemble non-$\chi$GOE  is therefore  expected to be a result of 
 the fact that
$\rho_{N \ \tx{corr.}}^{(1)}(x)$ precisely  cuts off the peaks  of $\rho^{(2)}_{N}(x)$,
and at the same time the 
term $\rho^{(4)}_{N \ \tx{corr.}}(x)$ must cause a highly oscillating $\rho_{N}^{(4)}(x)$.
When plotting $\rho_{N}^{(\beta)}(x)$ together with the associated 
$\rho_{N}^{(2)}(x)$ we indeed observe 
these highly non-trivial features. 
The results are illustrated in figure \ref{fig:6} and \ref{fig:7}. 

\begin{figure}[!htbp]
\begin{center}
\begin{tabular}{c c}
\includegraphics[scale=0.85]{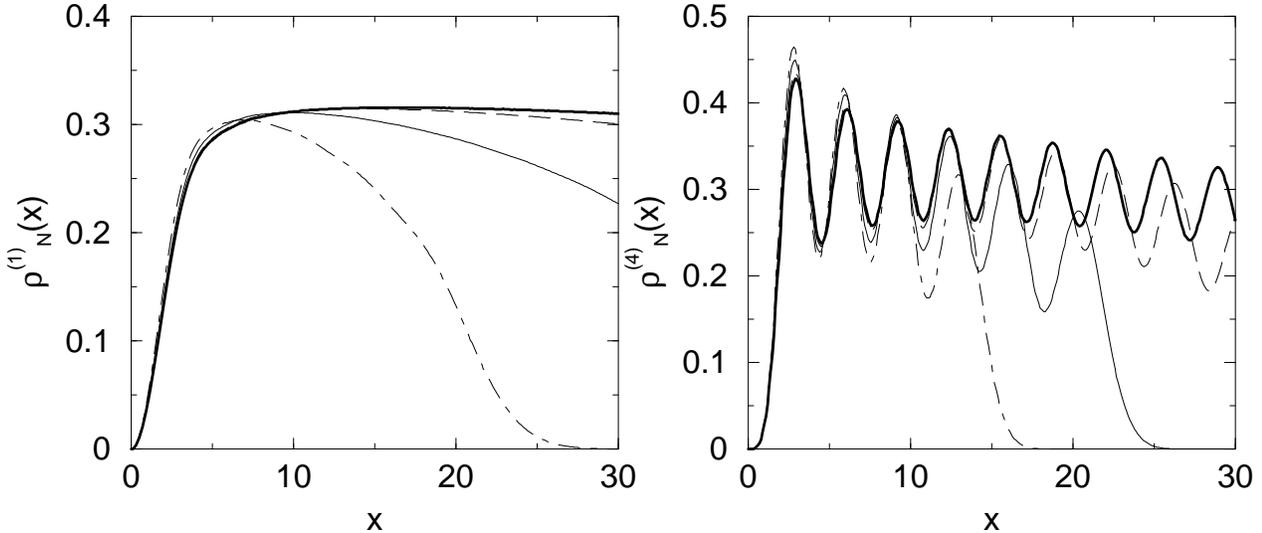} 
\end{tabular}
\end{center}
\caption{
The scaled spectral density $\rho_{N}^{(\beta)}(x)$,
for increasing $N$, 
 on the
left for $\beta=1$ and on the right for $\beta=4$. The convergens
in the vicinity of $x=0$ is clearly observed and we have
$\rho_{N}^{(\beta)}(x) \approx \rho_{s}^{(\beta)}(x)$ near $z=0$ as a consequence. In the limit $N \to \infty$ we have 
$\rho_{N }^{(\beta)}(x)\to \rho_{s}^{(\beta)}(x)$ and in the limit 
$x \to \infty$ we have  
$\rho_{s}^{(\beta)}(x) \to  \pi^{-1}$. }
\label{fig:8}
\end{figure}

\noi
In the limit $N \to \infty$ we have $\rho_{N }^{(\beta)}(x)\to
\rho_{s}^{(\beta)}(x)$, $\beta=1,4$.  This convergence, however, is
especially fast for $x$ close to zero.
This is illustrated in figure \ref{fig:8}, where we
have plotted $\rho_{N}^{(\beta)}(x)$, $\beta=1,4$, for different
values of $N$.  
The spectral density is an even function (see
(\ref{r-l})) and for finite $N$ we have the so-called half circles
(with a hole in the vicinity of $x=0$, due to the term $x^{N_{f}}$ in
the weight function), which is clearly seen in figure \ref{fig:8}. 
Because of the the Gaussian distribution the
microscopic spectral density fulfills $\rho_{s}(x)\to \pi^{-1}$ for $x
\to \infty$  and thus  the curves 
$\rho_{N}^{(\beta)}(x)$, $\beta=1,4$, are of order $\sim \pi^{-1}$
before the half circles start approaching zero (see also figure
\ref{fig:6} and \ref{fig:7}).

\subsection{The first spectral sum rule}
The spectral sum rules derived in field theory are all given by the microscopic spectral density $\rho_{s}$ of the Dirac operator \cite{3b,3k} 
and the result for $\rho_{s}$ derived from RMT can therefore be verified
through the knowledge of a  spectral sum rule.
The first sum involves the sum over the eigenvalues of the Dirac operator
in  second inverse power and in the limit $N \to \infty$ we have trivially  
\beq\label{sr}
\langle \sum_{\lambda_{k} \ne 0} 
\frac{1}{(N\Sigma\lambda_{k})^{2}} \rangle=
\int_{-\infty}^{\infty}dz \ \frac{\rho_{s}(z)}{z^{2}},
\eeq
where the sum is over all $\lambda_{k} \ne 0$ (both positive and
 negative) and the average is taken over all gauge fields.
For both field theories corresponding to the ensembles $\beta=1,4$ in
$(2+1)$ dimensions (see table \ref{treQCDteorier}) the same spectral
sum rule has been derived in \cite{3j,3j'} and it reads
\beq\label{snf} 
\langle
\sum_{\lambda_{k} \ne 0}\frac{1}{(N\Sigma\lambda_{k})^{2}} \rangle
=\frac{4}{N_{f}}.
\eeq 
According to this, the first spectral sum rule
involving the eigenvalues of the Dirac operator in two completely
different field theories in $(2+1)$ dimensions should be  identical.

\noi 
With the help of our numerical expressions for
$\rho_{s}^{(\beta)}(x)$ for the microscopic spectral density
$\rho_{s}^{(\beta)}(x)$ in the two ensembles $\beta=1,4$ we thus have
a consistency check on our results through (\ref{sr}) and (\ref{snf}).
From figure \ref{fig:7} and \ref{fig:8} it is easily seen that
the effect of the finite $N$ is very small for a relatively large $N$.
\begin{table}[h]
\begin{center}
\begin{tabular}{||c|c|c||c||}\hline\hline
$N_{f}$ &  $\int_{-\infty}^{\infty}dz \ \frac{\rho_{40}^{(1)}(z)}{z^{2}}$  &  
 $\int_{-\infty}^{\infty}dz \ \frac{\rho_{40}^{(4)}(z)}{z^{2}}$  & $\frac{4}{N_{f}}$  \\ \hline \hline
$2$       &  $0.402$         &  0.946        &  $2$ \\ \hline
$4$   & $0.148$  &  $0.436$   &  $1$ \\ \hline
$6$  & $0.092$  &  0.297  & $0.667$ \\ \hline 
$8$  & 0.067  & $0.227$ &  $0.500$        \\ \hline 
$12$ &0.044  & 0.155 & $0.333$\\ \hline \hline
\end{tabular}
\caption{The calculated first spectral sum rule, i.e the right hand side of (\ref{sr}),
in the two ensembles $\beta=1$ and $\beta=4$ together with the spectral sum rule 
(\ref{snf}).  }
\label{tabelsum}
\end{center}
\end{table}
As seen our results does not   agree with the spectral sum rule (\ref{snf}). 
In the ensemble $\beta=4$ 
 it seems the spectral sum rule approaches $2/N_{f}$ for large $N_{f}$. 
\\
Next we turn to the most elementary check of our results, which in
fact is an actual generation of matrices.

\section{The Monte Carlo simulation} 

We have performed a Monte Carlo simulation in order to numerically
verify the derived scaled spectral densities $\rho_{N}^{(\beta)}(x)$, $\beta=1,4$
The matrices in the generated ensemble are distributed according to (\ref{Z-chi}),
with all $m_{f}=0$.
 The simulations are done for
various values of the matrix size, $N$, and different values of $N_f$ in  the
two cases $\beta=1,4$.
\\
We diagonalize matrices from an ensemble made by a simple Metropolis
algorithm. The eigenvalues, $x$, are then collected to form a
histogram, $h(x)$, which is  rescaled
\begin{eqnarray}
x~& \rightarrow &~ x N ~\equiv~ z, \nonumber \\
h~& \rightarrow &~ \frac{h}{N}
\end{eqnarray}
to obtain $\frac{1}{N}h(\frac{z}{N})$. The rescaled histogram is then
compared to the plot of the analytically obtained scaled spectral
density.  Note that we compare the analytic plots for finite $N$ with
the corresponding simulation data. We thus expect the plots to match
for the whole spectrum.  For large $N$ the scaled spectral density
close to $z=0$ has the form of the microscopic spectral density. 
\\
We handle the two cases $\beta=1,4$  individually as follows.

\subsection{$\beta=1$}

\begin{figure}
\begin{center}
\includegraphics[scale=0.9]{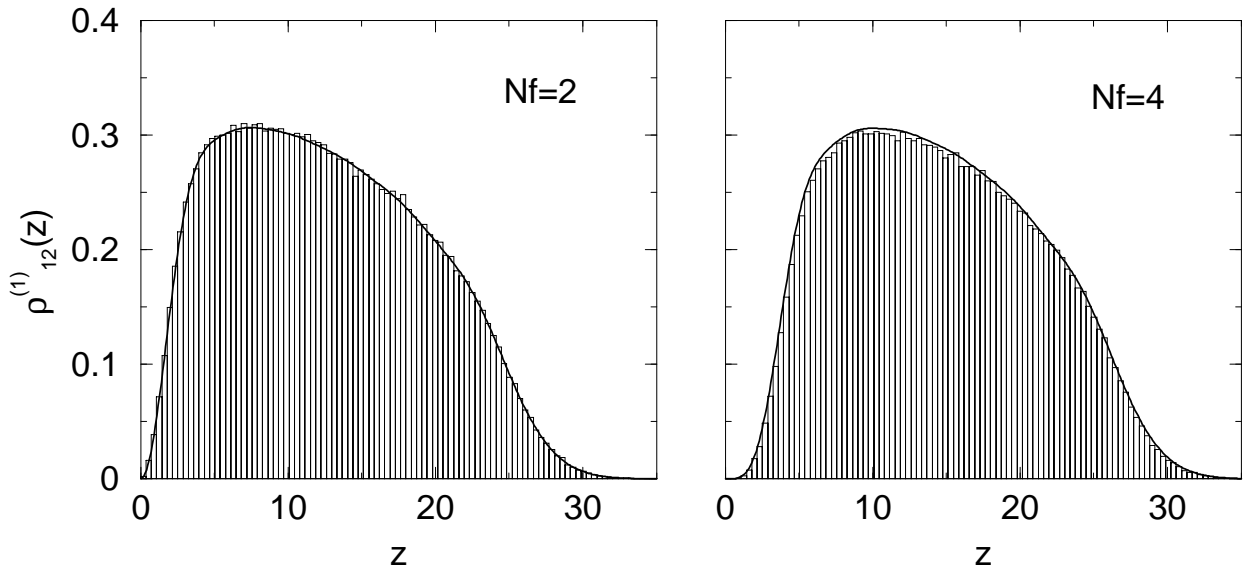}
\end{center}
\caption{Scaled spectral density of the non-$\chi$GOE ensemble, compared with the Monte Carlo data, $N=12$, $N_f=2$ and $4$.}
\label{fig:plot1}
\end{figure}

\begin{figure}
\begin{center}
\includegraphics[scale=0.9]{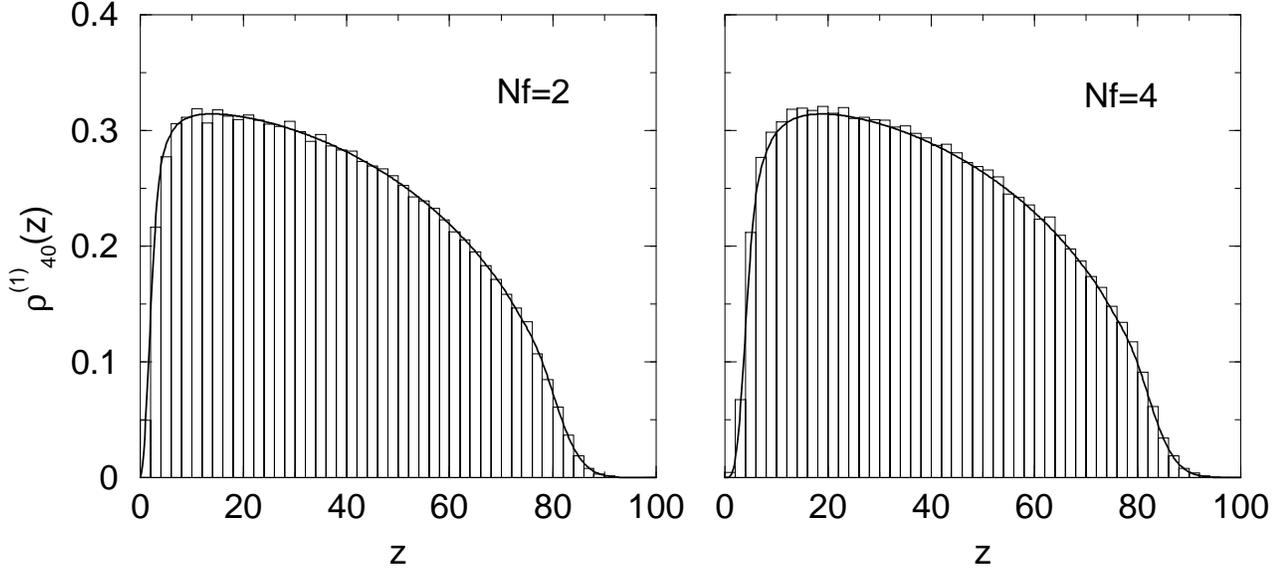}
\end{center}
\caption{Scaled spectral density of the non-$\chi$GOE ensemble, compared with the Monte Carlo data, $N=40$, $N_f=2$ and $4$.}
\label{fig:plot2}
\end{figure}

The Metropolis step consist in changing the entries in the matrix and
accepting these changes according to the action 
\beq
S(M)=\frac{N}{4}\tx{Tr}M^2 - N_f \ln \big(\det M \big).
\eeq
The changes is done in a way that explicitly maintain the symmetry of
the matrix ($M=M^T$).
\\
In this way we obtained an ensemble of 100,000 effectively
uncorrelated matrices for $N=12$ and 10,000 matrices for $N=40$, both
for $N_f=2$ and 4. The resulting histograms together with their
analytic equivalents can be seen in figure~\ref{fig:plot1} and
~\ref{fig:plot2}. We observe excellent agreement in all parts of
the spectrum, for all values of the parameters $N$ and $N_f$.

\subsection{$\beta=4$}
\begin{figure}
\begin{center}
\includegraphics[scale=0.9]{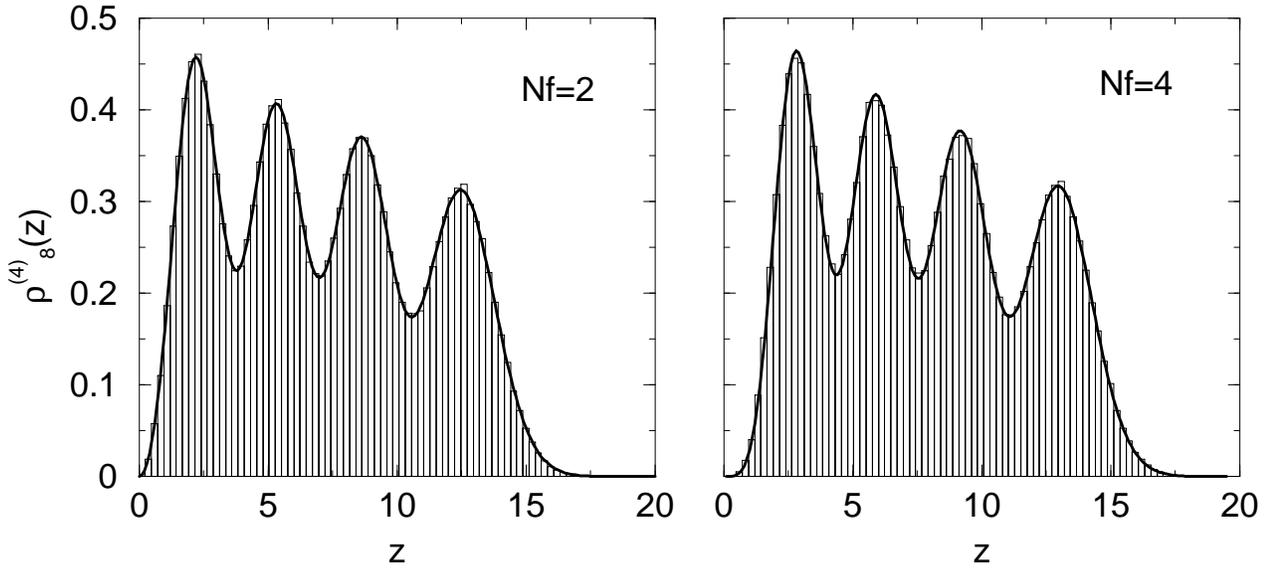}
\end{center}
\caption{Scaled spectral density of the non-$\chi$GSE ensemble, compared with the Monte Carlo data, $N=8$, $N_f=2,4$.}
\label{fig:plot3}
\end{figure}

\begin{figure}
\begin{center}
\includegraphics[scale=0.9]{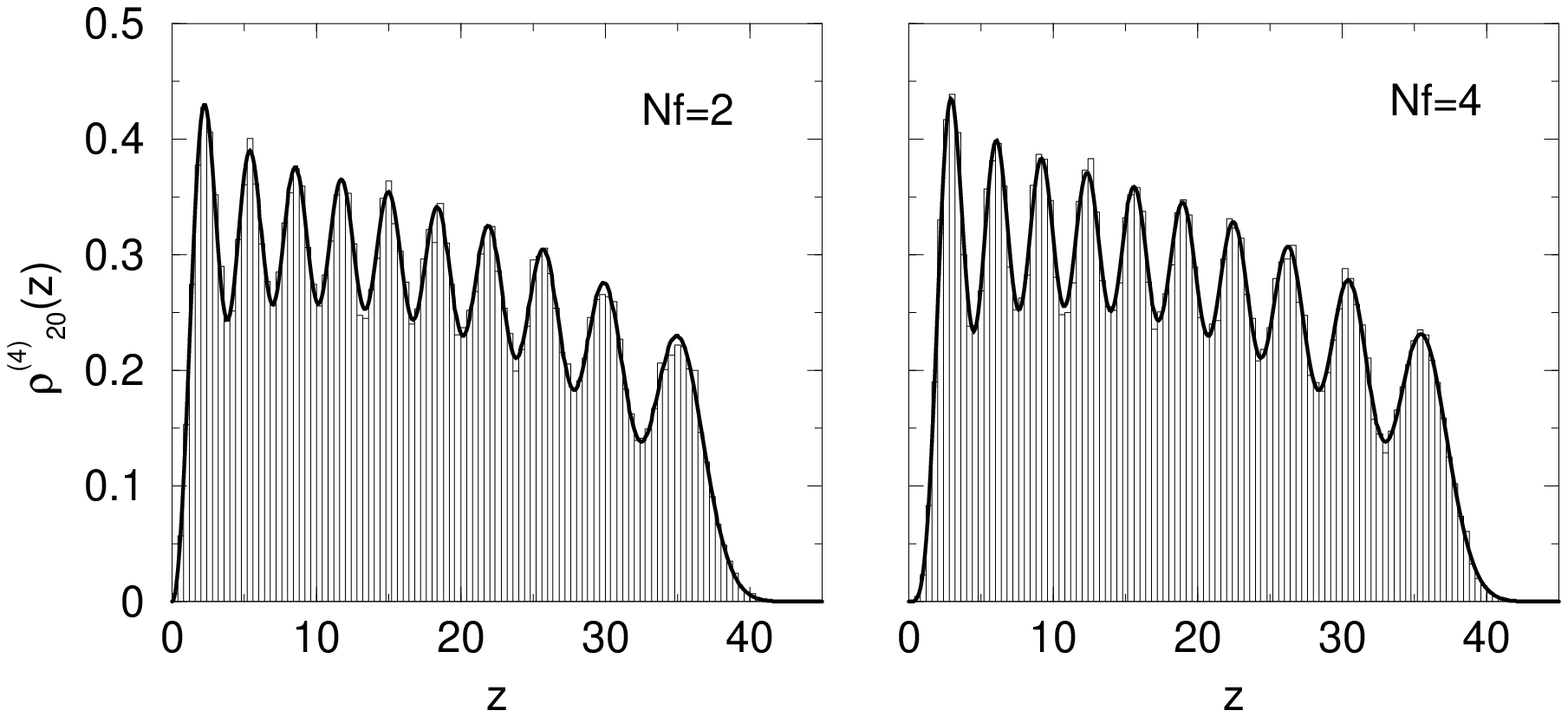}
\end{center}
\caption{Scaled spectral density of the non-$\chi$GSE ensemble, compared with the Monte Carlo data, $N=20$, $N_f=2,4$.}
\label{fig:plot4}
\end{figure}

\begin{figure}[!t]
\begin{center}
\includegraphics[scale=0.9]{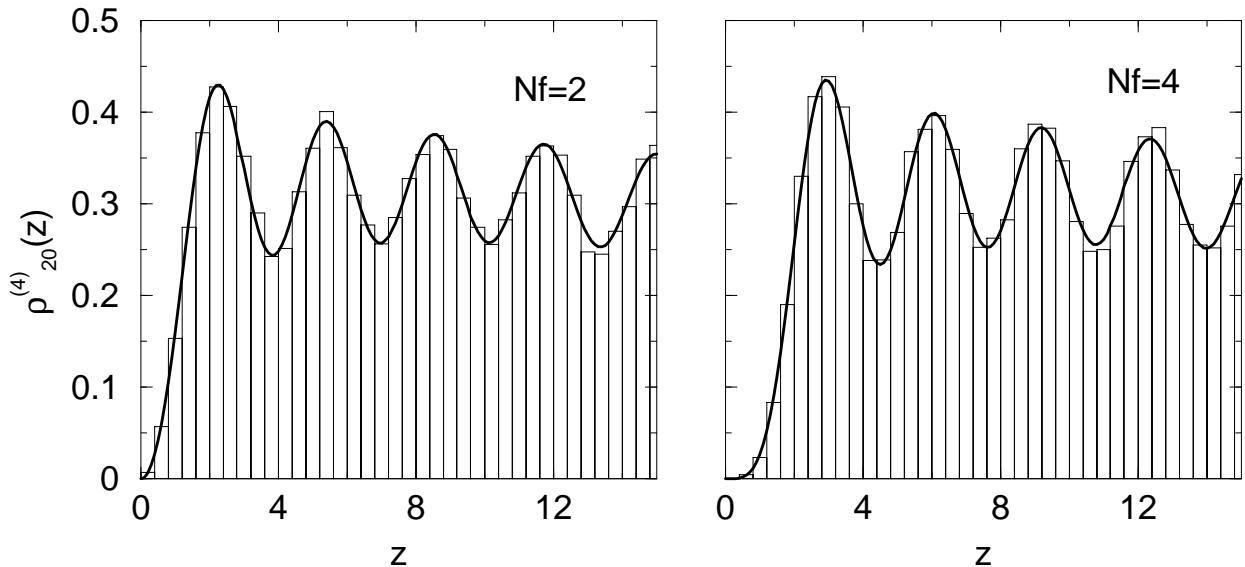}
\end{center}
\caption{Same plot as figure \ref{fig:plot4} but zoomed in on small values of $z$.}
\label{fig:zoom}
\end{figure}

In this case the Metropolis steps are accepted/rejected according to
\beq
S(M)=N\tx{Tr}M^2-N_f \ln \big(\det M \big)
\label{eq:GSE_action}
\eeq
A matrix from the symplectic ensemble can be represented
by a $2N\times 2N$ matrix, $M$. This matrix can be written as a sum of
direct-product matrices
\beq
M~=~M_0e_0 + M_1e_1 + M_2e_2 + M_3e_3
\eeq
where 
\begin{eqnarray}
e_0 & = & \left(\begin{array}{c c} 1 & 0 \\ 0 & 1 \end{array}\right) 
,~~~e_1  =  \left(\begin{array}{c c} 0 & i \\ i & 0 \end{array}\right) \\
e_2 & = & \left(\begin{array}{c c} 0 & -1 \\ 1 & 0 \end{array}\right) 
,~~~e_3  =  \left(\begin{array}{c c} i & 0 \\ 0 & -i \end{array}\right)
\end{eqnarray}
and the coefficients $M_1,M_2,M_3$ are real antisymmetric $N\times N$
matrices and $M_0$ is a real symmetric $N\times N$ matrix. In this way
the matrix $M$ has $2N$ eigenvalues which are doubly degenerate.
\\
The Metropolis update works on the four real matrices, $M_i$, and $M$
is then constructed from these and diagonalized. In this the way the
symmetry and structure of the matrices is kept at all times. For the
histogram only the $N$ different eigenvalues is used.
\\
When working in this $2N\times 2N$ representation the action,
(\ref{eq:GSE_action}), has to be changed. The trace gets twice as
large and the determinant is squared in this representation, so an
overall factor of 1/2 should be included:
\beq
S(M)=\frac{N}{2}\tx{Tr}M^2-\frac{N_f}{2}   \ln \big(\det  M
 \big)
\eeq
These simulations are about twice as slow, so the maximum matrix size
is half as big. We have done runs for $N=8$ and 20, with $N_f=2$ and
$4$, with the same statistics as in the $\beta=1$ case. The plots are
shown in figure~\ref{fig:plot3} and ~\ref{fig:plot4}. Here the
agreement is even more striking. Zooming in on the first few
eigenvalues in the $N=20$ case we see absolutely perfect agreement indeed, see
figure~\ref{fig:zoom}.

\section {Conclusions}

The purpose of this study has been to derive the kernels
$S_{N}^{(\beta)}(x,y)$, $\beta=1,4$, from which all spectral
correlation functions of the non-$\chi$Gaussian orthogonal
(non-$\chi$GOE) and symplectic (non-$\chi$GSE) ensembles can be
determined.  In the microscopic limit the two ensembles describes the 
microscopic spectral
correlations of the low energy Dirac operator in two Yang-Mills theory
in $(2+1)$ dimensions.  Thus the universality class non-$\chi$GOE
carries information about the spectral correlations in a $SU(2)$ gauge
theory with fundamental fermions, while the universality class
non-$\chi$GSE does the same for the $SU(N_{c})$ gauge theory, with
$N_{c}$ arbitrary, and fermions in the adjoint representation.  
\\
With the help of Widom's new method we have derived the kernels
$S_{N}^{(\beta)}(x,y)$, in non-$\chi$GOE ($\beta=1$) and in
non-$\chi$GSE ($\beta=4$) for massless fermions.  For non-$\chi$GOE
our result are valid for all integers $N_{f}$ and for non-$\chi$GSE we
have a result for even $N_{f}$.  We plotted the scaled spectral
density
$\rho_{N}^{(\beta)}(x)=N^{-1}S_{N}^{(\beta)}(N^{-1}x,N^{-1}x)$, for
different values of the parameters $N$ and $N_{f}$.  The plots of
$\rho_{N}^{(\beta)}(x)$ possesses the expected features, a flat
spectrum for $\beta=1$ and a highly oscillation spectrum for
$\beta=4$, and indeed we see perfect agreement with our computer
simulated spectra.  For large $N$ the scaled spectral density
$\rho_{N}^{(\beta)}(x)$ coincides with the microscopic spectral density
$\rho_{s}(x)$ in the vicinity of $x=0$.  Our calculated spectral
sum rules from $\rho_{N}^{(\beta)}(x)$ of course match the computer
generated ones perfectly. We have not found agreement
with the sum rules of \cite{3j,3j'}.
\\

\noindent
Recently the microscopic spectral densities of the massive non-$\chi$GOE and
non-$\chi$GSE have been derived by a different method
\cite{nagao_nishigaki}. In the massless limit these result seem to
match ours for large $N$.
\\
\\ 
\textbf{Acknowledgements }: Poul
Henrik Damgaard is thanked for many fruitful discussions. The work of
C. Hilmoine is partially supported by a Rosenfeld Fellowship.

\appendix
\section{Interchange of integration and the microscopic limit}

In this Appendix we look at the large $n$-behavior of two types of
general integrals involving generalized Laguerre
polynomials. Specifically we examine when interchange of integration
and the microscopic limit is allowed. To this end Lebesgue's Majorant
Theorem is used. The first section is relevant for the calculation of
$\varepsilon \psi_{i}$ in the limit $N \to \infty$, see
(\ref{epsiiullim}). The result of section 2 is relevant for the
calculation of the elements of the matrix $B$ in the limit $N \to
\infty$, see (\ref{h20}).

\subsection{$\displaystyle\lim_{n \to \infty} \int_{0}^{\infty}dy \ n^{(1+a)-\alpha}  L_{n}^{\alpha}(2y) y^{a}  \exp(-y)$}
\label{apa1}
We define
\begin{equation}\label{}
\mathcal{I}^{\alpha,a}_{n}  \equiv \int_{0}^{\infty}dy \  n^{(1+a)-\alpha} L_{n}^{\alpha}(2y) y^{a}  \exp(-y),
\end{equation}
where $n$ is a positive integer, $a>-1$ and $\alpha$ are real numbers.
We wish to calculate
\begin{equation}\label{1appen2}
\lim_{n \to \infty} \mathcal{I}^{\alpha,a}_{n}=
\lim_{n \to \infty} \int_{0}^{\infty}dy \ n^{(1+a)-\alpha}  L_{n}^{\alpha}(2y) y^{a}  \exp(-y),
\end{equation}
and therefore  examine when the interchange of the limit and the
integration is allowed. In order to do this we employ Lebesgue's
Majorant Theorem. At first we convince ourselfs that the sequence of functions in the integrand is convergent. The integral in (\ref{1appen2}) is rewritten by the substitution $u=(2ny)^{1/2}$, and we get
\begin{equation}\label{1appen3}
\lim_{n \to \infty} \mathcal{I}^{\alpha,a}_{n}=\lim_{n \to \infty} \int_{0}^{\infty}du \ n^{(1+a)-\alpha}  L_{n}^{\alpha}(\frac{u^2}{n}) u^{2a+1} n^{-a-1}2^{-a} \exp(-\frac{u^2}{2n}).
\end{equation}
The integrand in (\ref{1appen3})
\beq\label{fn1}
f_{n}(u)\equiv
n^{(1+a)-\alpha} L_{n}^{\alpha}(\frac{u^2}{n}) u^{2a+1} n^{-a-1}2^{-a} \exp(-\frac{u^2}{2n}),
\eeq
is convergent for $n \to \infty$ : Inserting the well-known asymptotic formula 
\cite{high}
\begin{equation}\label{LimL}
\lim_{n \to \infty}\frac{1}{n^{\alpha}}L^{\alpha}_{n}(\frac{x^2}{n})=x^{-\alpha} J_{\alpha}(2x),
\end{equation}
we get
\begin{equation}\label{a1}
\lim_{n \to  \infty}f_{n}(u)=
\lim_{n \to  \infty} n^{(1+a)-\alpha} L_{n}^{\alpha}(\frac{u^2}{n}) u^{2a+1} n^{-a-1}2^{-a} \exp(-\frac{u^2}{2n})=2^{-a}u^{2a+1-\alpha}J_{\alpha}(2u).
\end{equation}
Thus, the integrand in (\ref{1appen2}) is convergent. 
We must now look for an integrable majorant $\mathcal{M}(y)$,
fulfilling for all $n$ 
\begin{equation}\label{numint}
\vert f_{n}(y) \vert=
\vert n^{(1+a)-\alpha}  L_{n}^{\alpha}(2y) y^{a}  \exp(-y)  \vert<\mathcal{M}(y).
\end{equation}
If we can find such a function $\mathcal{M}(y)$, the Lebesgue Theorem states that interchange of the limit and integration is allowed in (\ref{1appen2}). 
In this case we can insert (\ref{a1}) in (\ref{1appen2}).
\\
We start by splitting up the integral into integrals, in the first
 integrating over the interval $]0,\epsilon[$
 for some finite $\epsilon$
 and in the second over
$]\epsilon, \infty[$.
Since the first integral is finite for every $n$
 we focus only on the second integral.
 By choosing a sufficiently
large $n'$ we can  use an asymptotic
expression for the integrand. We then check if the absolute value of
this expression has an integrable majorant for all $n>n'$, see
(\ref{numint}). Using the asymptotic property (valid for $x>0$) \cite{magnus}
\begin{equation}\label{asymL2}
L^{\alpha}_{n}(x)=\frac{1}{\pi} e^{\frac{x}{2}}
x^{-(\frac{2\alpha-1}{4})} n^{(\frac{2\alpha-1}{4})}
\cos\big[2(nx)^{\frac{1}{2}}-\frac{\alpha \pi}{2}-\frac{\pi}{4}\big]+\mathcal{O}( n^{\frac{2\alpha-3
}{4}}).
\end{equation}
for $L_{n}^{\alpha}(2y)$ in (\ref{1appen2}), we get
\begin{eqnarray}\label{a2} 
\lefteqn{\lim_{n \to \infty} \mathcal{I}^{\alpha,a}_{n}} & {}& \nonumber\\ 
 &\approx&\lim_{n \to \infty} \int_{\epsilon}^{\infty}dy \ n^{(1+a-\alpha)} \nonumber\\
&{}& 
\times
\Big\{ \pi^{-1/2} \exp(y) y^{-(\frac{2\alpha+1}{4})}2^{-(\frac{2\alpha+1}{4})}n^{(\frac{2\alpha-1}{4})} \cos\Big[2(2ny)^{\frac{1}{2}}-\frac{\alpha \pi}{2}-\frac{\pi}{4}\Big]\Big\} y^a \exp(-y) \nonumber \\
&=&
\pi^{-1/2}2^{-(\frac{2\alpha+1}{4})} \lim_{n \to \infty}  n^{(1+a-\frac{\alpha}{2}-\frac{1}{4})} \int_{\epsilon}^{\infty}dy \ y^{-\frac{2\alpha+1}{4}+a} \cos\Big[2(2ny)^{\frac{1}{2}}-\frac{\alpha \pi}{2}-\frac{\pi}{4}\Big], 
\end{eqnarray}
for  $n >n'$.
\\
We know that the sequence of functions in (\ref{a2}) is convergent, and the function
\begin{equation}\label{m}
\mathcal{M}(y)=c\  \pi^{-1/2}2^{-(\frac{2\alpha+1}{4})} \ y^{-\frac{2\alpha+1}{4}+a}, \qquad c>1,
\end{equation}
is then a majorant to the absolute value of the integrand in (\ref{1appen2}) for $n>n'$ and some $c >1$,
which take account of the term $\mathcal{O}$ in (\ref{asymL2}).
Since we need $\int_{\epsilon}^{\infty} dy\ \mathcal{M}(y)< \infty $ we must have $-(\alpha/2+1/4)+a<-1$.
\\
So we conclude, \emph{if}
\beq\label{beta}
\alpha/2-a>3/4,
\eeq
\emph{then} (\ref{m}) 
\emph{is a integrable  majorant and we can then interchange the order of the limit and the integral in (\ref{1appen2})}. 
Asumming this in (\ref{a1}), gives us
\begin{equation} 
\lim_{n \to \infty} \mathcal{I}^{\alpha,a}_{n}=2^{-a}\int_{0}^{\infty}du \ u^{2a+1-\alpha}J_{\alpha}(2u),
\end{equation}
for $\alpha/2-a>3/4$.
\\
On the contrary, if $\alpha /2-a \le 3/4$, then the integrand in
(\ref{1appen2}) has no majorant, $\mathcal{M}(y)$, which fulfills
$\int_{0}^{\infty}dy \mathcal{M}(y)<\infty$.
This follows from the fact that the integral from $\epsilon$ to infinite of the absolut value of the
integrand in (\ref{a2}) is not finite.
Substituting $z=2ny$ in (\ref{a2}) gives an integrand on the form
$
z^{-\delta}\cos(2z^{1/2}+k),
$  
where $\delta=-(\alpha/2+1/4)+a$ (the $n$ dependence vanishes of course).
The smallest majorant to 
$
\vert z^{-\delta}\cos(2z^{1/2}+k) \vert,
$
 is 
$
C\vert  z^{-\delta}\cos(2z^{1/2}+k) \vert,
$
for some $C>1$.
This majorant is not  integrable when $\delta=-(\alpha/2+1/4)+a \ge -1$,
and 
thus we have no  integrable majorant.
When the absolut value of the asymptotic function does not integrate to a finite number, then
of course no majorant exist for which the integral is finite
(see (\ref{numint})).\\
\emph{We conclude, when }
\beq\label{betb}
\alpha/2-a \le 3/4,
\eeq
\emph{interchange of integration and the limit in} (\ref{1appen2}) \emph{is not legal}.
In this case  
we first have to solve the integral in (\ref{1appen2}) and subsequently
derive an expression in the limit $n \to \infty$.

\subsection{
$\displaystyle \lim_{n \to \infty}  \int_{0}^{\infty}dt \
 n^{\lambda} L_{n}^{\alpha}(2t) t^{\beta}  \exp(-t) \int_{0}^{t}du \ L_{n}^{\bar{\alpha}}(2u) u^{\bar{\beta}}  \exp(-u)$}
We will now look at the  limit $n \to \infty$ of integrals of the following type
\begin{equation}\label{LimImn}
\mathcal{I}^{\alpha, \beta, \bar{\alpha},\bar{\beta}}_{n}
\equiv n^{\lambda} \int_{0}^{\infty}dt \ L_{n}^{\alpha}(2t) t^{\beta}  \exp(-t) \int_{0}^{t}du \ L_{n}^{\bar{\alpha}}(2u) u^{\bar{\beta}}  \exp(-u),
\end{equation}
where $\alpha,\bar{\alpha},\beta,\bar{\beta}>-1$ and $\lambda$ is real
and $n$ is a positive integer.
The matrix $B$ in section 5 is given by integrals of this type, and we are interested in an
 expression for $B$ in the limit $n \to\infty$.
It is assumed that the integrand is convergent for $n \to\infty$, meaning that $\lambda$ has a certain
value, the size of which is  irrelevant for the question addressed in this Appendix.
\\
Analogous to the previous section we investigate whether the limit and the integration can be interchanged in the expression
\begin{equation}\label{2appen1}
\lim_{n \to \infty}\mathcal{I}^{\alpha,\beta,\bar{\alpha},\bar{\beta}}_{n}  =
\lim_{n \to \infty}  \int_{0}^{\infty}dt \
 n^{\lambda} L_{n}^{\alpha}(2t) t^{\beta}  \exp(-t) \int_{0}^{t}du \ L_{n}^{\bar{\alpha}}(2u) u^{\bar{\beta}}  \exp(-u).
\end{equation}
By assumtion the sequence   
\begin{equation}\label{kint}
f_{n}(t)\equiv  n^{\lambda} L_{n}^{\alpha}(2t) t^{\beta}  \exp(-t) \int_{0}^{t}du \ L_{n}^{\bar{\alpha}}(2u) u^{\bar{\beta}}  \exp(-u),
\end{equation}
is convergent ( compare with (\ref{fn1}) og (\ref{a1}) ).
We must determine whether or not the 
 absolut value  of the  integrandt in (\ref{2appen1}),
that is $\vert f_{n}\vert $,
has an integrabel majorant $\mathcal{M}(t)$.  
Like in 
Appendix B.1 we persuit this question by choosing a large $n$ and
 put in the asymptotic relation (\ref{asymL2}) for both
 Laguerre polynomials :
\begin{eqnarray}\label{2a2} 
\lefteqn{\lim_{n \to \infty} \mathcal{I}^{\alpha,\beta,\bar{\alpha,\bar{\beta}}}_{n} 
\approx \pi^{-1}2^{-\frac{\alpha+\bar{\alpha}+1}{2}} 
\lim_{n \to \infty}  n^{\lambda} \quad \times}
\nonumber\\
& &
 \int_{\epsilon}^{\infty}dt \ \vert \ n^{\frac{\alpha}{2}-\frac{1}{4}}
 \ t^{-(\frac{\alpha}{2}+\frac{1}{4}-\beta)}\cos\Big[2(2nt)^{\frac{1}{2}}-\frac{\alpha \pi}{2}-\frac{\pi}{4}\Big] 
\int_{\epsilon}^{t} du \ 
 n^{\frac{\bar{\alpha}}{2}-\frac{1}{4}} u^{-(\frac{\bar{\alpha}}{2}+\frac{1}{4}-\bar{\beta})}  
\cos\Big[2(2nu)^{\frac{1}{2}}-\frac{\bar{\alpha} \pi}{2}-\frac{\pi}{4}\Big]\ \vert.
\nonumber\\
 \end{eqnarray}
Like in Appendix B.1 we skip all contributions involving  irrelevant integrals over $]0,\epsilon[$, because in
these cases the exact integrals are finite. 
The  sequence in (\ref{2appen1}), that is $f_{n}$ above, is convergent by assumtion.
Thus for all $n$ the absolut value of the 
integrand in (\ref{2a2}) is always smaller than the function
\begin{equation}\label{maj}
c \ t^{-(\frac{\alpha}{2}+\frac{1}{4}-\beta)}
\int_{\epsilon}^{t}du \ u^{-(\frac{\bar{\alpha}}{2}+\frac{1}{4}-\bar{\beta})},
\end{equation}
where $c>1$. 
Letting $\delta=-(\frac{\alpha}{2}+\frac{1}{4}-\beta)$ and $\rho=-(\frac{\bar{\alpha}}{2}+\frac{1}{4}-\bar{\beta})$
we have trivially
\begin{equation}\label{in}
c \ \int_{\epsilon}^{\infty}dt  \ t^{\delta}
\int_{\epsilon}^{t}du \ u^{\rho} =
\frac{c}{\rho+1} \int_{\epsilon}^{\infty}dt \
\Big(t^{\rho +\delta+1}-t^{\delta}\epsilon^{\rho+1}\Big).
 \end{equation}
We see immediately that for
\begin{equation}\label{bet1}
\rho+\delta+1=-\frac{\alpha+\bar{\alpha}-1}{2}+\beta+\bar{\beta} <-1
\quad \textrm{and} \quad \delta=-(\frac{\alpha}{2}+\frac{1}{4}-\beta)<-1
\end{equation}
the integral in
 (\ref{in}) is finite.
For large $n$
we therefore  have that the absolute value of the integrand in (\ref{kint}) always is smaller than the function
(\ref{maj}),
and if 
(\ref{bet1}) is valid the function (\ref{maj}) is an integrable majorant.
\\
On the contrary we see from (\ref{2a2}) in case of 
$\delta=-(\frac{\alpha}{2}+\frac{1}{4}-\beta) \ge-1$ \tx{and}
$\rho=-(\frac{\bar{\alpha}}{2}+\frac{1}{4}-\bar{\beta})\ge-1$,
then the 
integral over  $t$  is \emph{divergent}. 
It follows that the absolut value of the integrand in (\ref{2appen1})
does not have an integrable majorant, since the asymptotic expression does not even have one.
\\
The other cases of values of $\rho$ and
$\delta$ are not relevant for us and we ignore them.
\\
\\
Summarizing the discussion we have
\label{ombyt}
\begin{itemize}
 \item[(1)] If (\ref{bet1}) is valid, then
 the absolut value of  the  integrand in (\ref{2appen1}) has an integrable  majorant (\ref{maj}).
Thus interchange of the limit and  integration in (\ref{2appen1}) is legal:
\begin{equation}\label{ombyt+}
\lim_{n \to \infty}\mathcal{I}^{\alpha,\beta,\bar{\alpha,\bar{\beta}}}_{n}
= \int_{0}^{\infty}dt \ \lim_{n \to \infty}  n^{\lambda} L_{n}^{\alpha}(2t) t^{\beta}  \exp(-t) \int_{0}^{t}du \ L_{n}^{\bar{\alpha}}(2u) u^{\bar{\beta}}  \exp(-u).
\end{equation}
\item[(2)]
If $-(\frac{\alpha}{2}+\frac{1}{4}-\beta) \ge -1 $ \emph{and}
$-(\frac{\bar{\alpha}}{2}+\frac{1}{4}-\bar{\beta}) \ge -1$, then
 the absolut value of  the  integrand in 
 (\ref{2appen1})
has no majorant   $\mathcal{M}(t)$ fulfilling $\int_{0}^{\infty}dt \ \mathcal{M}(t)<\infty$. 
Therefore interchange  
of the limit and the integration in
 (\ref{2appen1}) is not 
allowed.  
In this case we must solve the integral before the limit is taken.
\end{itemize}

\section{Two integrals solved}

In this Appendix we find expressions for two different types of integrals.
We wish to derive an expression for the function
\beq\label{eL}
\mathcal{E}_{[\bar{\alpha},\bar{\beta},n]}(x) \equiv
\varepsilon \
 L_{n}^{\bar{\alpha}}(x^{2})
x^{\bar{\beta}} \exp(-\frac{x^{2}}{2})= 
 \int_{-\infty}^{\infty}dy \  \varepsilon(x-y)
 L_{n}^{\bar{\alpha}}(y^{2})
y^{\bar{\beta}} \exp(-\frac{y^{2}}{2}), 
\eeq
where $\bar{\alpha}>-1$ is a real number and $n$, $\bar{\beta}$ are integers.

For integers $m$ and $n$ we calculate the following numbers
\begin{eqnarray}\label{B}
\mathcal{B}_{ij}&\equiv&
 \big(
\varepsilon \ L_{n}^{\bar{\alpha}}(x^{2})x^{\bar{\beta}} \exp(-\frac{x^{2}}{2})
\ , \
 L_{m}^{\alpha}(x^{2}) x^{\beta}  e^{-\frac{x^{2}}{2}}
 \big)=
\int_{-\infty}^{\infty} dx \
L_{m}^{\alpha}(x^{2}) x^{\beta}  e^{-\frac{x^{2}}{2}}
\varepsilon   \
 L_{n}^{\bar{\alpha}}(x^{2})
x^{\bar{\beta}} e^{-\frac{y^{2}}{2}} \nonumber\\
& = & \int_{-\infty}^{\infty}dx \
 L_{m}^{\alpha}(x^{2})
x^{\beta}e^{ -\frac{x^{2}}{2} }
 \int_{-\infty}^{\infty}dy \   \varepsilon(x-y)
 L_{n}^{\bar{\alpha}}(y^{2})
y^{\bar{\beta}} e^{-\frac{y^{2}}{2}}, 
\end{eqnarray}
The index $i$ refers to $m,\alpha$, and $\beta$, while $j$ refers to $n,\bar{\alpha}$ and $\bar{\beta}$. $m$, $n$, $\beta$ and $\bar{\beta}$ are integers and $\alpha$
and $\bar{\alpha}$ are real numbers greater than -1. 
\\
The function  (\ref{eL}) is a part of the inner product in 
$\mathcal{B}_{ij}$ and first we therefore derive an expression for (\ref{eL}).

\subsection{$\displaystyle \varepsilon \
 L_{n}^{\bar{\alpha}}(x^{2})
x^{\bar{\beta}} \exp(-\frac{x^{2}}{2}) $}
We start by defining
\beq
E(\lambda,x)\equiv \varepsilon
 x^{\lambda}\Big(
x e^{-\frac{x^{2}}{2}}\Big),
\eeq
where $\lambda \ge -1$ is an integer.
We have the following recursion formula
\begin{eqnarray}\label{Erec}
E(\lambda,x)&=&
 \int_{-\infty}^{\infty}dy \ \varepsilon (x-y)
 y^{\lambda}\Big(
y e^{-\frac{y^{2}}{2}}\Big)  \nonumber\\
& = &\frac{1}{2}\Big[\int_{-\infty}^{x}dy-\int_{x}^{\infty}dy \Big]
 y^{\lambda}\Big(
y e^{-\frac{y^{2}}{2}}\Big)  \nonumber\\
&=& -[\frac{1}{2}e^{-\frac{y^{2}}{2}}  y^{\lambda}]_{-\infty}^{x}
+[\frac{1}{2}e^{-\frac{y^{2}}{2}}  y^{\lambda}]_{x}^{\infty}
+\int_{-\infty}^{\infty} dy \ \varepsilon(x-y)
\lambda y^{\lambda-1}\Big(e^{-\frac{y^{2}}{2}}\Big)\nonumber\\
 &= &
-e^{-\frac{x^{2}}{2}}x^{\lambda}
+ \lambda \int_{-\infty}^{\infty}dy \ \varepsilon(x-y)\
 y^{\lambda-2}\Big(y e^{-\frac{y^{2}}{2}}\Big)  \nonumber\\
&= &-e^{-\frac{x^{2}}{2}}x^{\lambda}+\lambda E(\lambda-2,x), 
\end{eqnarray}
for $\lambda \ge 1$.
For $\lambda =-1$ we have
\begin{eqnarray}\label{-1}
E(-1,x)&=& \int_{-\infty}^{\infty}dy \ \varepsilon (x-y)
 e^{-\frac{y^{2}}{2}} \nonumber\\
 & =&
\frac{1}{2}\Big\{ \int_{-\infty}^{-x}dy +\int_{-x}^{x}dy -\int_{x}^{\infty}\Big\} \ 
 e^{-\frac{y^{2}}{2}} \nonumber\\
 & =&
 \int_{0}^{x}dy \ e^{-\frac{x^{2}}{2}}=
  \sqrt{\frac{\pi}{2}}
  \textrm{erf}(\frac{x}{\sqrt{2}}).
\end{eqnarray}
And for $\lambda=0$ we get
\begin{eqnarray}\label{0}
E(0,x)&=& \int_{-\infty}^{\infty}dy \ \varepsilon (x-y)
 ye^{-\frac{y^{2}}{2}}=-e^{-\frac{x^{2}}{2}}.
\end{eqnarray}
For $\lambda$ \emph{even} we immediately get from (\ref{Erec}) and (\ref{0})
\beq\label{llige}
E(\lambda,x)=-e^{-\frac{x^{2}}{2}}\Big(x^{\lambda}+\lambda x^{\lambda-2}+\dots+
\lambda (\lambda-2)\cdots 4 \cdot 2   \Big).
\eeq
In the case of \emph{odd} $\lambda$, (\ref{Erec}) and (\ref{-1}) give us :
\beq\label{lulige}
E(\lambda,x)=-e^{-\frac{x^{2}}{2}} \Big(x^{\lambda} +
  \lambda x^{\lambda-2}+\dots+
 \lambda (\lambda-2)\cdots 5 \cdot 3 x \Big)+
\lambda(\lambda-2)\cdots 5 \cdot 3
    \sqrt{\frac{\pi}{2}}
  \textrm{erf}(\frac{x}{\sqrt{2}}).
\eeq
Writing out the terms of the  Laguerre polynomials through \cite{magnus}
\beqa\label{Lsum}
L^{\alpha}_{n}(x)=\sum_{m=0}^{n}(-1)^{m}
{n+\alpha \choose n-m }
\frac{x^{m}}{m!}=
\sum_{m=0}^{n}(-1)^{m}
\frac{\Gamma(n+\alpha+1)}{\Gamma(n-m+1)\Gamma(m+\alpha+1)}  \frac{x^{m}}{m!},
\eeqa
we get
\beqa\label{eLsum}
\varepsilon \
 L_{n}^{\bar{\alpha}}(x^{2})
x^{\bar{\beta}} \exp(-\frac{x^{2}}{2})
  \nonumber
 & =&
\int_{-\infty}^{\infty}dy \  \varepsilon(x-y)
 L_{n}^{\bar{\alpha}}(y^{2})
y^{\bar{\beta}}e^{ -\frac{y^{2}}{2} } \nonumber\\
&=& \sum_{i=0}^{n} a_{i}\int_{-\infty}^{\infty}dy
\ \varepsilon(x-y)
 \ y^{2i+\bar{\beta}-1}
\Big(ye^{-\frac{y^{2}}{2}}  \Big) \nonumber\\
&=& a_{0}\int_{-\infty}^{\infty}dy\ \varepsilon(x-y)  \ y^{\bar{\beta}-1}
\Big(ye^{-\frac{y^{2}}{2}}  \Big) +
 a_{1}\int_{-\infty}^{\infty}dy\ \varepsilon(x-y)  \ y^{\bar{\beta}+1}
\Big(ye^{-\frac{y^{2}}{2}}  \Big)+\nonumber\\
& & \quad \quad      \quad \dots +
 a_{n}\int_{-\infty}^{\infty}dy\ \varepsilon(x-y)  \ y^{2n+\bar{\beta}-1}
\Big(ye^{-\frac{y^{2}}{2}}  \Big),
\eeqa
where $a_{i}$ are the coefficients to $x^{i}$ in (\ref{Lsum}).
\\
Assuming that $\bar{\beta}$ is \emph{odd}, all powers in (\ref{eLsum})
are even and we use (\ref{llige}) on each term to give us
\beqa\label{eLalige}
\lefteqn{
\varepsilon \
 L_{n}^{\bar{\alpha}}(x^{2})
x^{\bar{\beta}} \exp(-\frac{x^{2}}{2})
}\nonumber\\
&=&
-e^{-\frac{x^{2}}{2}}\Big[
 a_{0}\Big(x^{\bar{\beta}-1}+(\bar{\beta}-1)x^{\bar{\beta}-3}+\dots+
 (\bar{\beta}-1)(\bar{\beta}-3)\cdots 4 \cdot 2   \Big)+ \nonumber\\
& & \qquad \quad a_{1}\Big(x^{\bar{\beta}+1}+(\bar{\beta}+1)x^{\bar{\beta}-1}+\dots+
 (\bar{\beta}+1)(\bar{\beta}-1)\cdots 4 \cdot 2   \Big)+\cdots \cdots \nonumber\\
& &\qquad \quad + a_{n}\Big(x^{\bar{\beta}+2n-1}+(\bar{\beta}+2n-1)x^{\bar{\beta}-3}+\dots+
 (\bar{\beta}+2n-1)(\bar{\beta}+2n-3)\cdots 4 \cdot 2   \Big)
 \Big]\nonumber\\
&=&
-e^{-\frac{x^{2}}{2}}
\Big[
\Big\{
a_{0}+
(\bar{\beta}+1)a_{1}+(\bar{\beta}+3)(\bar{\beta}+1)a_{2}+\dots+
(\bar{\beta}+2n-1)(\bar{\beta}+2n-3) \cdots (\bar{\beta}+1)a_{n}
\Big\}\times \nonumber\\
& & \qquad \qquad \Big(x^{\bar{\beta}-1}+ (\bar{\beta}-1)x^{\bar{\beta}-3}+
 (\bar{\beta}-1) (\bar{\beta}-3)x^{\bar{\beta}-5}+\cdots +
  (\bar{\beta}-1) (\bar{\beta}-3)\cdots 4 \cdot 2
\Big)
\nonumber\\
&& \qquad +x^{\bar{\beta}+1}
\Big(a_{1}+
(\bar{\beta}+3)a_{2}+(\bar{\beta}+5)  (\bar{\beta}+3)a_{3}+\dots+
(\bar{\beta}+2n-1)(\bar{\beta}+2n-3) \cdots (\bar{\beta}+3)a_{n}
\Big)+\nonumber\\
& & \qquad \qquad
x^{\bar{\beta}+3}
\Big(a_{2}+
(\bar{\beta}+5)a_{3}+\dots+
(\bar{\beta}+2n-1) \cdots (\bar{\beta}+5)a_{n}
\Big)+ \dots \nonumber\\
& &\qquad \qquad \qquad \qquad \qquad \qquad  \dots+
x^{\bar{\beta}+2n-1}a_{n}\Big].\nonumber\\
\eeqa
\newpage
When $\bar{\beta}$ is \emph{even} we use (\ref{lulige}) in (\ref{eLsum}), and we get
\beqa\label{eLaulige}
\lefteqn{
\varepsilon \
 L_{n}^{\bar{\alpha}}(x^{2})
x^{\bar{\beta}} \exp(-\frac{x^{2}}{2})
}\nonumber\\
&=& 
-e^{-\frac{x^{2}}{2}}\Big[
 a_{0}\Big(x^{\bar{\beta}-1}+(\bar{\beta}-1)x^{\bar{\beta}-3}+\dots+
 (\bar{\beta}-1)(\bar{\beta}-3)\cdots 5 \cdot 3 \ \textrm{erf}\big(\frac{x}{\sqrt{2}} \big)  \Big)+ \nonumber\\
& &\qquad \quad a_{1}\Big(x^{\bar{\beta}+1}+(\bar{\beta}+1)x^{\bar{\beta}-1}+\dots+
 (\bar{\beta}+1)(\bar{\beta}-1)\cdots  5 \cdot 3 \ \textrm{erf}\big(\frac{x}{\sqrt{2}}  \Big)+\cdots \cdots \nonumber\\
& &\quad + a_{n}\Big(x^{\bar{\beta}+2n-1}+(\bar{\beta}+2n-1)x^{\bar{\beta}-3}+\dots+
 (\bar{\beta}+2n-1)(\bar{\beta}+2n-3)\cdots  5 \cdot 3 \ \textrm{erf}\big(\frac{x}{\sqrt{2}}   \Big)
 \Big].\nonumber\\
& &
\eeqa
Using the definitions \cite{high}
\beqa\label{!!1}
(2k)!! \equiv 2k(2k-2)\cdots4\cdot \ 2=2^{k}\ \Gamma(k+1),
\eeqa
\beqa\label{!!2}
(2k-1)!! \equiv (2k-1)(2k-3)\cdots 3 \cdot   \ 1
=\pi^{-\frac{1}{2}}\ 2^{k} \ \Gamma(k+\frac{1}{2}),
\eeqa
where $k$ is a positive integer, we can write (\ref{eLalige}) and
(\ref{eLaulige}) in a more compact way.  
\\
For every $a_{i}$, $0\le i \le
n$, the righthand side of eq. (\ref{eLalige}) is equal to $-e^{-x^{2}/2}$ times a 
polynomial of order $(\bar{\beta}+2i-1)$. This polynomial can be rewritten as 
\beq\label{khu}
a_{i}\sum_{j=0}^{\frac{\bar{\beta}-1}{2}+i}x^{\bar{\beta}-1-2j+2i} \ 2^{j}\
\frac{\Gamma\big(\frac{\bar{\beta}+1}{2}+i\big)}{\Gamma\big(\frac{\bar{\beta}+1}{2}+i-j\big)}.
\eeq
A similar expression exist for the polynomial in the framed part of (\ref{eLaulige}).
Using these expressions for the polynomial parts of 
(\ref{eLaulige}) and 
(\ref{eLalige}) and
inserting 
the Laguerre coefficients $a_{i}$ of $x^{i}$ in (\ref{Lsum})
in (\ref{eLaulige}) and 
(\ref{eLalige}) leads to 
\\
\begin{tabular}{|p{18cm}|}
\hline
\beqa\label{eLulige}
& &\mathcal{E}_{[\bar{\alpha}, \bar{\beta},n]}(x)=
\varepsilon \
 L_{n}^{\bar{\alpha}}(x^{2})
x^{\bar{\beta}} \exp(-\frac{x^{2}}{2})=   \nonumber\\
& &-e^{\frac{-x^{2}}{2}} \Big[
\sum_{i=0}^{n}
\frac{(-1)^{i}}{\Gamma(i+1)}\frac{\Gamma(n+1+\bar{\alpha})}{\Gamma(n+1-i)\Gamma(\bar{\alpha}+1+i)}  
\ \sum_{j=0}^{\frac{\bar{\beta}-1}{2}+i\ \{-\frac{3}{2}\}} x^{(\bar{\beta}-1-2j+2i)} \  2^{j}
\frac{\Gamma(\frac{\bar{\beta}+1}{2}+i)}{\Gamma(\frac{\bar{\beta}+1}{2}+i-j)}\Big]\nonumber\\
 & &\Big\{+
\sum_{i=0}^{n}
\frac{(-1)^{i}}{\Gamma(i+1)}\frac{\Gamma(n+1+\bar{\alpha})}{\Gamma(n+1-i)\Gamma(\bar{\alpha}+1+i)}  2^{  (i+\frac{\bar{\beta}}{2})}  
  \Gamma\Big(i+\frac{\bar{\beta}+1}{2}\Big)
\Big[-e^{-\frac{x^{2}}{2}} \frac{x}{ {\sqrt{\pi}  }} \ 
  +
\sqrt{\frac{1}{2}} \ \textrm{erf}\big(\frac{x}{\sqrt{2}}\big)\Big]\Big\}.\nonumber\\
\eeqa
\n
\\
\hline
\end{tabular}
\n
\n
For \emph{even} $\bar{\beta}$ we must include the terms in the brackets $\{\dots \}$, while these are neglected for \emph{odd} $\bar{\beta}$. 
For $\bar{\beta}=0$ a  modification of the solution with the brackets is valid :
for $i=0$ we must delete the term $-xe^{-x^{2}/2}/\sqrt{\pi}$.   
All contributions from a negative upper limit in the summation are set to zero 
(for instance for $\bar{\beta}=2$ and $i=0$ the term involving  $ \sum_{j=0}^{-1}$
equals zero).

\subsection{$\displaystyle
\int_{-\infty}^{\infty} dx \
L_{m}^{\alpha}(x^{2}) x^{\beta}  e^{-\frac{x^{2}}{2}}
\varepsilon   \
 L_{n}^{\bar{\alpha}}(x^{2})
x^{\bar{\beta}} e^{-\frac{x^{2}}{2}} $}

 Using (\ref{eLulige})
we are now able to calculate (\ref{B}).
Since $\beta$, $\bar{\beta}$  are integers, the interior and exterior functions in (\ref{B}) 
$$f_{int}(x)\equiv L_{n}^{\bar{\alpha}}(x^{2})x^{\bar{\beta}} e^{-\frac{x^{2}}{2}}$$ 
and 
$$f_{ext}(x) \equiv L_{m}^{\alpha}(x^{2}) x^{\beta}
e^{-\frac{x^{2}}{2}} $$
are either even or odd. In this case it is easily shown that 
$\mathcal{B}_{ij}$ is non-zero only when $\bar{\beta}$ is even and $\beta$ is odd or vice visa.
In the following we assumed that $\beta$ and $\bar{\beta}$ fulfills
this.  Using the relation $\mathcal{B}_{ij}=-\mathcal{B}_{ji}$ we can
make sure that the function in the interior integral in
(\ref{B}) is \emph{odd}.
If $\bar{\beta}$ is \emph{odd} this is automatically the case and we have
\beq
\mathcal{B}_{ij}=\int_{-\infty}^{\infty}dx \ f_{ext}(x) \varepsilon f_{int}(x),
\eeq
with $\varepsilon f_{int}(y)$ given by (\ref{eLulige}).
In case of  $\bar{\beta}$ \emph{even} we simply use $\mathcal{B}_{ij}=-\mathcal{B}_{ji}$ to switch the two functions
\beq\label{byd}
\mathcal{B}_{ij}=-\mathcal{B}_{ji}=
-\int_{-\infty}^{\infty}dx \ f_{int}(x) \varepsilon f_{ext}(x).
\eeq
where now $\varepsilon f_{ext}(x)$ is given by (\ref{eLulige}).
\\  
So from now on we assume $\bar{\beta}$ is \emph{odd} and $\beta$ \emph{even}.
\\
Denoting by $p_{\bar{\beta}-1+2n}$ the $(\bar{\beta}-1+2n)$ order polynomial from (\ref{eLulige}), we want to calculate
\begin{eqnarray}\label{Bbarbul}
\mathcal{B}_{ij}&=&\int_{-\infty}^{\infty} dx \
L_{m}^{\alpha}(x^{2}) x^{\beta}  e^{-x^{2}}p_{\bar{\beta}-1+2n}(x)
\end{eqnarray}
Since both $p_{\bar{\beta}-1+2n}$ and $L_{m}^{\alpha}(x^{2}) x^{\beta}  e^{-\frac{x^{2}}{2}}$  are \emph{even}, we have
\begin{eqnarray}\label{}
\mathcal{B}_{ij}&=&-2\int_{0}^{\infty} dx \
L_{m}^{\alpha}(x^{2}) x^{\beta}  e^{-x^{2}}
p_{\frac{\bar{\beta}-1+2n}{2}}(x^{2}).
\end{eqnarray}
Making the substitution $z=x^{2}$, gives
\begin{eqnarray}\label{her20}
\mathcal{B}_{ij}&=&-\int_{0}^{\infty} dz \
L_{m}^{\alpha}(z) z^{\frac{\beta-1}{2}}  e^{-z}
    p_{\bar{\beta}-1+2n}(z^{\frac{1}{2}}).
\end{eqnarray}
Let us rewrite $p_{\bar{\beta}-1+2n}$ using (\ref{eLalige})
\beqa\label{her20'}
p_{\bar{\beta}-1+2n}(z^{\frac{1}{2}})&=&
b_{\bar{\beta}+2n-1  } z^{\frac{\bar{\beta}+2n-1}{2}}+
b_{(\bar{\beta}+2n-3)} \ z^{\frac{\bar{\beta}+2n-3}{2}}+\dots+
b_{(\bar{\beta}+3)} \ z^{\frac{\bar{\beta}+3}{2}}+
b_{(\bar{\beta+1})}\ z^{\frac{\bar{\beta}+1}{2}}+
\nonumber\\
& &T \Big(
z^{\frac{\bar{\beta}-1}{2}}+b_{(\bar{\beta}-3)}z^{\frac{\bar{\beta}-3}{2}}+
b_{(\bar{\beta}-5)}z^{\frac{\bar{\beta}-5}{2}}+\dots+b_{0}
 \Big)
\eeqa
The coefficients $b_{i}$ are given by products of $\bar{\beta}$ and
$a_{i}$, and $T$ is the factor in brackets ($\{\dots \}$) in (\ref{eLalige}).
Using (\ref{!!1}) and (\ref{!!2})
$T$ can be reduced to :
\beqa\label{T}
T&=&
\Big\{
a_{0}+
(\bar{\beta}+1)a_{1}+(\bar{\beta}+3)(\bar{\beta}+1)a_{2}+\dots+
(\bar{\beta}+2n-1)(\bar{\beta}+2n-3) \cdots (\bar{\beta}+1)a_{n}
\Big\}\nonumber\\
&=&
\sum_{i=0}^{n}(-1)^{i} \ 2^{i}\frac{\Gamma(\frac{\bar{\beta}+1}{2}+i)}{\Gamma(\frac{\bar{\beta}+1}{2})\Gamma(i+1)}
\frac{\Gamma(n+1+\bar{\alpha})}{\Gamma(n+1-i)\Gamma(\bar{\alpha}+1+i)}. 
\eeqa
By insertion of (\ref{her20'}) in (\ref{her20}) we get
\begin{eqnarray}\label{her19}
\mathcal{B}_{ij}&=&-
\int_{0}^{\infty} dz \
L_{m}^{\alpha}(z)   e^{-z} z^{\frac{\bar{\beta}+\beta}{2}}
\Big[ b_{(\bar{\beta}+2n-1 ) }z^{n-1}
+b_{(\bar{\beta}+2n-3)}  z^{n-2}+
\dots
+b_{(\bar{\beta}+3)}
z
+b_{(\bar{\beta}+1)}
\Big]\nonumber\\
& & \qquad \qquad
-T\int_{0}^{\infty} dz \
L_{m}^{\alpha}(z)   e^{-z} z^{\frac{\bar{\beta}-1}{2}}
 \Big(
z^{\frac{(\bar{\beta}-1)}{2}}+b_{(\bar{\beta}-3)}z^{\frac{\bar{\beta}-3}{2}}+
b_{(\bar{\beta}-5)}z^{\frac{\bar{\beta}-5}{2}}+\dots+b_{0}
 \Big).
\end{eqnarray}
We restrict the parameters to the following four cases
\beqa\label{bg1}
\alpha=\frac{\bar{\beta}+\beta}{2},  \qquad \textrm{and} \quad (n-1)<m \quad \textrm{or}\quad (n-1)=m, 
\eeqa
\beqa\label{bg2}
\alpha=\frac{\bar{\beta}+\beta}{2}+ 1, \qquad \textrm{and} \quad (n-1)<m \quad \textrm{or}\ (n-1)=m.
\eeqa
In case (\ref{bg1}), orthonormality reduces (\ref{her19}) to
\begin{eqnarray}\label{her18}
\mathcal{B}_{ij}&=&
-T\int_{0}^{\infty} dz \
L_{m}^{\alpha}(z)   e^{-z} z^{\frac{\bar{\beta}-1}{2}}
 \Big(
z^{\frac{\bar{\beta}-1}{2}}+b_{(\bar{\beta}-3)}z^{\frac{\bar{\beta}-3}{2}}+
b_{(\bar{\beta}-5)}z^{\frac{\bar{\beta}-5}{2}}+\dots+b_{0}
 \Big),
\end{eqnarray}
for  $(n-1)<m$, while for $(n-1)=m$ we have
\begin{eqnarray}\label{her17}
\mathcal{B}_{ij}&=&
-
 b_{(\bar{\beta}+2n-1 ) }
\int_{0}^{\infty} dz \
L_{m}^{\alpha}(z)   e^{-z} z^{\alpha}
z^{m}
-\nonumber\\
& & \qquad \quad
T\int_{0}^{\infty} dz \
L_{m}^{\alpha}(z)   e^{-z} z^{\alpha}
 \Big(
z^{\frac{\bar{\beta}-1}{2}}+b_{(\bar{\beta}-3)}z^{\frac{\bar{\beta}-3}{2}}+
b_{(\bar{\beta}-5)}z^{\frac{\bar{\beta}-5}{2}}+\dots+b_{0}
 \Big).\
\end{eqnarray}
The first term in (\ref{her18}) reduces to 
\beqa
- b_{(\bar{\beta}+2n-1 ) }
\int_{0}^{\infty} dz \
L_{m}^{\alpha}(z)   e^{-z} z^{\alpha}
z^{m}=
- b_{(\bar{\beta}+2n-1 ) }
\int_{0}^{\infty} dz \
L_{m}^{\alpha}(z)   e^{-z} z^{\alpha}
\Big(\frac{m!}{(-1)^{m}}L_{m}^{\alpha}(z) \Big)\nonumber\\
=- b_{(\bar{\beta}+2n-1 ) }  \frac{m!}{(-1)^{m}}h_{m}^{\alpha}=
- b_{(\bar{\beta}+2n-1 ) }  \frac{m!}{(-1)^{m}}\frac{\Gamma(\alpha+1+m)}{m!}\nonumber\\
=  - b_{(\bar{\beta}+2n-1 ) }(-1)^{m}\Gamma(\alpha+1+m)
=-a_{n}(-1)^{m}\Gamma(\alpha+1+m) \nonumber\\
=-\frac{(-1)^{n}}{n!} (-1)^{m}\Gamma(\alpha+1+m)
=-\frac{(-1)^{m+1}}{(m+1)!} (-1)^{m}\Gamma(\alpha+1+m) =\frac{\Gamma(\alpha+m+1)}{\Gamma(m+2)}.
\eeqa
Working out explicitly what the $b_i$'s are in terms of $\Gamma$-functions and using \cite{high}
\beqa\label{intL2}
\int_{0}^{\infty}dx \ e^{-x} x^{\gamma-1}L_{n}^{\mu}(x)=
\frac{\Gamma[\gamma]\Gamma[1+\mu+n-\gamma]}{\Gamma[n+1]\Gamma[1+\mu-\gamma]}.
\eeqa
on each term in (\ref{her18}) gives us the result
\n
\n
\begin{tabular}{|p{17cm}|}
\hline
\beqa\label{bereg1B}
 \mathcal{B}_{ij}&=& 
-\sum_{i=0}^{n}(-1)^{i} \ 2^{i}\frac{\Gamma(\frac{\bar{\beta}+1}{2}+i)}{\Gamma(\frac{\bar{\beta}+1}{2})\Gamma(i+1)}
\frac{\Gamma(n+1+\bar{\alpha})}{\Gamma(n+1-i)\Gamma(\bar{\alpha}+1+i)} 
\ \times  \nonumber\\ 
& &
\qquad \qquad \qquad
\frac{1}{\Gamma(m+1)}
\sum_{j=0}^{\frac{\bar{\beta}-1}{2}}
\frac{\Gamma(\alpha-j)\Gamma(1+m+j)}{\Gamma(1+j)}
\  2^{j}
\frac{\Gamma(\frac{\bar{\beta}+1}{2})}{\Gamma(\frac{\bar{\beta}+1}{2}-j)}(+)
\frac{\Gamma(\alpha+m+1)}{\Gamma(m+2)},\nonumber\\
\eeqa
for
\beq\label{pmb1}
\frac{(\bar{\beta}+\beta)}{2}=\alpha,~~~(n-1)\leq m,~~\textrm{and}~~ \bar{\beta}\ \textrm{odd}, 
\eeq
The last term is only added in the case $(n-1)=m$.
\n
\\
\hline
\end{tabular}
\n
\n

In the case (\ref{bg2}), only one term in the first line of (\ref{her19}) survives because of orthonormality. It reads
\beqa
- b_{(\bar{\beta}+1)}
\int_{0}^{\infty} dz \
L_{m}^{\alpha}(z)   e^{-z} z^{\frac{\bar{\beta}+\beta}{2}}
&=&-
b_{(\bar{\beta}+1)}\int_{0}^{\infty} dz \
L_{m}^{\alpha}(z)   e^{-z} z^{\frac{\bar{\beta}+\beta}{2}+1}
z^{-1}\nonumber\\
&=&  -
b_{(\bar{\beta}+1)}\int_{0}^{\infty} dz \
L_{m}^{\alpha}(z)   e^{-z} z^{\alpha}
z^{-1}
=-b_{(\bar{\beta}+1)}\Gamma(\alpha).
\eeqa
By (\ref{eLalige}) and (\ref{her20'}) we have
\beqa
 b_{(\bar{\beta}+1)}&=&\Big(a_{1}+
(\bar{\beta}+3)a_{2}+(\bar{\beta}+5)  (\bar{\beta}+3)a_{3}+\dots+
(\bar{\beta}+2n-1)(\bar{\beta}+2n-3) \cdots (\bar{\beta}+3)a_{n}.
\Big)
\nonumber\\
&=&
\frac{T-a_{0}}{(\bar{\beta}+1)}.
\eeqa
The contribution from the second line in (\ref{her19}) is given by (\ref{bereg1B}) without the $(+)$ term and with the substitution $\alpha \to \alpha-1$. 
Collecting the parts we get
that  
for \\
\\

\begin{tabular}{|p{17cm}|}
\hline
\beq\label{pmb}
\frac{\bar{\beta}+\beta}{2} + 1=\alpha, ~~~
\ (n-1) \le m ~~~\textrm{, and}~~~ \bar{\beta}\ \textrm{odd}
\eeq
we have
\beq\label{bereg2}
 \mathcal{B}_{ij}=\Big[\textrm{Righthand side of eq. (\ref{bereg1B})
with $\alpha \to \alpha-1$ and \emph{without} the term $(+)$ }\Big] +\frac{T-a_0}{\bar{\beta}+1}\ \Gamma(\alpha),
\eeq
where $T$ is given by eq. (\ref{T}).
\\ \\
\hline
\end{tabular}


\end{document}